\input epsf.tex    

\documentstyle[11pt]{article}

\newcommand{\bmat}{\left(\begin{array}}
\newcommand{\emat}{\end{array}\right)}

\def\preal{{\rm Re\,}}
\def\pim{{\rm Im\,}}

\def\yzero{\smash{\hbox{$y\kern-4pt\raise1pt\hbox{${}^\circ$}$}}}

\def\a{\alpha}
\def\b{\beta}
\def\g{\gamma}

\def\beq{\begin{equation}}
\def\eeq{\end{equation}}
\def\beqa{\begin{eqnarray}}
\def\eeqa{\end{eqnarray}}

\def\om{\omega}

\def\-{\hphantom{-}}
\def\ov{\overline}

\def\s2{\frac{1}{2}}

\def\oh{\frac{1}{2}}
\def\beq{\begin{equation}}
\def\eeq{\end{equation}}
\def\beqa{\begin{eqnarray}}
\def\eeqa{\end{eqnarray}}

\def\IF{\relax{\rm I\kern-.18em F}}
\def\II{\relax{\rm I\kern-.18em I}}

\def\Hh{{\cal H}}
\def\Bb{{\cal B}}

\def\cp{{\cal P}}
\def\IC{\bf C}
\def\IZ{\bf Z}
\def\IR{\bf R}

\def\IS{\bf S}
\def\IP{\bf P}

\def\z2z2{$\IC^3/(\IZ_2\times\IZ_2)$}

\def\NN{{\cal N}}

\def\e{\epsilon}

\def\Dsl{\,\raise.15ex\hbox{/}\mkern-13.5mu D} %this one can be subscripted

 \def\cp#1{\relax\ifmmode {\IP\kern-2pt{}_{#1}}\else $\IP\kern-2pt{}_{#1}$\=fi}
\newcommand{\drawsquare}[2]{\hbox{%
\rule{#2pt}{#1pt}\hskip-#2pt%  left vertical
\rule{#1pt}{#2pt}\hskip-#1pt%  lower horizontal
\rule[#1pt]{#1pt}{#2pt}}\rule[#1pt]{#2pt}{#2pt}\hskip-#2pt%  upper horizontal
\rule{#2pt}{#1pt}}% right vertical

\newcommand{\fund}{\raisebox{-.5pt}{\drawsquare{6.5}{0.4}}}%  fund
\newcommand{\antifund}{\overline{\fund}}

\begin{document}

%----------------------------------------------------------------------%
%  numbering equations with section number
%----------------------------------------------------------------------%
\makeatletter \@addtoreset{equation}{section} \makeatother
\renewcommand{\theequation}{\thesection.\arabic{equation}}
%----------------------------------------------------------------------%
%  title page
%----------------------------------------------------------------------%  

\pagestyle{empty}
\vspace*{.5in}
\rightline{FTUAM-04/13, IFT-UAM/CSIC-04-33}
\rightline{\tt hep-th/0407132}
\vspace{1.5cm}

\begin{center}
\LARGE{\bf Branes on Generalized Calibrated Submanifolds} \\[10mm]

\medskip

\large{Juan F. G. Cascales, Angel M. Uranga} \\
{\normalsize {\em Instituto de F\'{\i}sica Te\'orica, C-XVI \\
Universidad Aut\'onoma de Madrid \\
Cantoblanco, 28049 Madrid, Spain \\
{\tt juan.garcia@uam.es, angel.uranga@uam.es} \\[2mm]}}

\end{center}

\smallskip

\begin{center}
\begin{minipage}[h]{14.5cm}
{\small
We extend previous results on generalized calibrations to describe
supersymmetric branes in supergravity backgrounds with diverse fields 
turned on, and provide several new classes of examples. As an important 
application, we show that supersymmetric D-branes in compactifications 
with field strength fluxes, and on $SU(3)$-structure spaces, wrap 
generalized calibrated submanifolds, defined by simple conditions in terms 
of the underlying globally defined, but non-closed, 2- and 3-forms. We 
provide examples where the geometric moduli of D-branes (for instance 
D7-branes in 3-form flux configurations) are lifted by the generalized 
calibration condition. In addition, we describe supersymmetric D6-branes 
on generalized calibrated 3-submanifolds of half-flat manifolds, which 
provide the mirror of B-type D-branes in IIB  CY compactifications with 
3-form fluxes. Supersymmetric sets of such D-branes carrying no homology 
charges are mirror to supersymmetric sets of D-branes which are 
homologically non-trivial, but trivial in K-theory. As an additional
application, we describe models with chiral gauge sectors, realized in
terms of generalized calibrated brane box configurations of NS- and 
D5-branes, which are supersymmetric but carry no charges, so that no 
orientifold planes are required in the compactification.}
\end{minipage}
\end{center}

\newpage                                                        

%----------------------------------------------------------------------%
%  Resetting of counters
%----------------------------------------------------------------------%
\setcounter{page}{1} \pagestyle{plain}
\renewcommand{\thefootnote}{\arabic{footnote}}
\setcounter{footnote}{0}
%----------------------------------------------------------------------%
%  Paper begins
%----------------------------------------------------------------------%

\section{Introduction}

Recently it is becoming manifest that the dynamics of string theory is 
particularly rich and interesting in the presence of backgrounds with 
non-trivial field strength fluxes (or their dual versions). One example 
of such situations is provided by the gauge/gravity correspondence 
\cite{AdSCFT}, where the additional ingredients break conformal 
invariance and (partially) supersymmetry. Another situation of this kind 
is in compactifications of string/M - theory with field strength fluxes 
for $p$-form fields (see e.g. \cite{fluxes,gukov,drs,gkp,kst}), or in 
supersymmetric compactifications on non-Calabi-Yau manifolds of $SU(3)$ 
structure (see e.g. \cite{kstt,glmw,GranaMinasian}).

These compactifications show interesting properties, like moduli 
stabilization, warped geometries, and tractable supersymmetry breaking. 
Thus they represent an important step in constructing phenomenologically 
appealing string vacua (see \cite{blt,cascur,cgqu} for 
explicit model building with semirealistic gauge sectors), and in 
understanding string vacua with generically few moduli.

An important question in this framework is the behaviour of D-branes in 
such backgrounds. There are several levels at which this question may be 
addressed. For instance, at the topological level, there are additional 
consistency conditions for the possibilities of brane wrapping in the 
presence of fluxes \cite{baryons,mms}. Or, from the world-volume perspective, 
the fluxes may induce new terms in the action for world-volume fields (see 
e.g. \cite{softterms,soft2,ciu2}). In the present paper, we center on the 
similarly important aspect of characterizing supersymmetry preserving 
branes in such backgrounds.

In supersymmetric cases, these backgrounds are generalizations of 
flat space configurations or of compactification on special holonomy 
manifolds. In these simpler cases, supersymmetric branes are associated 
to calibrated submanifolds, which are volume minimizing. Thus, we may expect 
that supersymmetric branes in generalized supersymmetric backgrounds are
associated to generalized calibrations. These have been studied in particular
simple cases in \cite{gpt,NS5s} and more extensively discussed in \cite{martelli}, see also \cite{dallagata}. They can be described as closed forms which, once
restricted to the brane world-volume, provide not just the volume of the 
wrapped submanifold, but also contributions from additional coupling of the 
branes to the background. Alternatively, they contain a non-closed piece 
reproducing the volume of the wrapped submanifold, and which can be 
complemented by additional terms to a closed calibration. Hence branes on 
generalized calibrated submanifolds minimize not their volume, but rather 
their action.

In the present paper we describe D-branes in generalized calibrated 
submanifolds in supersymmetric backgrounds including non-trivial profiles 
for the metric, dilaton, and diverse $p$-form fields. Our analysis and 
examples extend previous results on generalized 
calibrations in \cite{gpt,NS5s,martelli,dallagata} to cases with additional backgrounds (and also those obtained in 
\cite{noforce} from the D$p$-brane worldvolume perspective). 

We moreover argue that the description in terms of generalized calibrations
allows to characterize supersymmetric wrapped branes in compactifications with
field strength fluxes and on non-Calabi-Yau $SU(3)$-structure manifolds\footnote{The interplay between G-structures and generalized calibrations has been discussed in \cite{martelli}}. We 
show that generalized calibrated branes can also be characterized by imposing
conditions of the familiar kind on the natural bispinor forms, namely the 
3-form $\Omega$, and a complexified version of the 2-form $J$, which are
globally defined, but not closed. Thus, generalized calibrations provide the
natural setup in which to describe supersymmetric branes in flux
compactifications.

Finally, we exploit a further property of branes on generalized calibrated 
submanifolds, namely that they are stable and supersymmetric even though 
they may not carry any topological charge. This allows new possibilities 
for model building, by using compactifications with branes but no 
orientifold planes. We illustrate this possibility in a compactification 
including brane-box configurations of NS5-branes and D5-branes on 
generalized calibrated submanifolds, leading to a chiral gauge sector.

\medskip

The paper is organized as follows. In section \ref{calibrations} we 
review calibrations and their relation to supersymmetry, and derive 
generalized calibrations from supersymmetry algebra considerations. In 
section \ref{examples} we present examples of D-branes on generalized 
calibrated submanifolds in backgrounds involving non-trivial metrics, 
dilaton profiles, and field strength fluxes. These include and extend 
previous examples in the literature. 

In section \ref{warmup} we introduce a warm-up example, and describe 
D3-branes in 3-form flux backgrounds, and a T-dual version of D4-branes 
in non-Calabi-Yau metric backgrounds. We show they correspond to 
generalized calibrations, and that this is crucial for them to be 
supersymmetric, since they turn out to carry no ($\IZ$-valued) charge. In 
section \ref{general} we propose a description of generalized calibrated 
submanifolds in compactifications with fluxes, or on $SU(3)$-structure 
manifolds, in terms of simple conditions on the globally defined 2- and 
3-forms. We provide explicit examples of branes in 3-form flux backgrounds, 
and on their mirror half-flat manifolds, supporting the proposal. In 
addition we show that in certain cases the condition to be generalized 
calibrated fixes some of the moduli of the wrapped submanifolds. For 
D7-branes, this encodes the fact that supersymmetric fluxes make D7-brane 
geometric moduli massive, a fact of relevance in the proposal in 
\cite{kklt} to stabilize K\"ahler moduli via gauge non-perturbative 
effects on D7-branes. 

In section \ref{building} we illustrate new model building possibilities with
generalized calibrated branes, by describing a model with 4d chiral gauge 
sectors from generalized calibrated brane box configurations. Finally, 
section \ref{fcomments} contains some final remarks. We list some 
conventions in the appendix.

\section{Calibrations and generalized calibrations}
\label{calibrations}

\subsection{Standard Calibrations}

In this section, we work in a pure metric background with no other fields 
turned on. Let us start by reviewing the formal definition of calibration
\cite{HarveyLaw} \footnote{For a nice recent review and broader 
bibliography, see \cite{Joyce} and references therein.}.

Let $M$ be a $d$-dimensional Riemannian manifold and $\varphi$ be a 
$k$-form on $M$, $k<d$. The form $\varphi$ is said to be a calibration on 
$M$ if: \textbf{i)} d$\varphi$=0 ($\varphi$ is closed) and \textbf{ii)} 
for any tangent $k$-plane to $M$, $\xi$, it satisfies $\varphi|_\xi \le 
vol_\xi$, where $\varphi|_\xi$ denotes the pullback of $\varphi$ on 
$\xi$, and $vol_\xi$ is the induced volume form on $\xi$. This pullback 
will typically be $\varphi|_\xi = \a\cdot vol_\xi$ for $\a\in \IR$ and so 
$\varphi|_\xi \le vol_\xi$ when $\a \le 1$.

A $k$-dimensional submanifold of $M$, $N$, is said to be calibrated 
by $\varphi$ if at any point $x \in N$, it satisfies $\varphi |_{T_xN}=
vol_{T_xN}$, i.e. if condition (ii) of the calibration is saturated at 
each point of the submanifold. It is hence intuitively clear that such 
calibrated submanifold will be volume minimizing in its topological 
class, as stated by the theorem of Harvey and Lawson \cite{HarveyLaw}. 

Calibrations are useful in string theory because they realize BPS-like
conditions, and so allow to classify supersymmetric extended objects of the 
background. This can be derived from the supersymmetry algebra as 
follows. The presence of an extended object in a generic supergravity 
background leads to a deformation of the superalgebra, that should now 
include an appropriate central charge, see e.g. \cite{Azcarraga,Mfromsuper}. 
Schematically, the anticommutator of supercharges in the presence of an 
extended $p$-dimensional object is \footnote{Our notation is as follows. 
We use $M$, $N$, $\ldots$ for ambient (generally curved) spacetime 
indices, and $A,B,...$ for tangent space indices. For the world-volume of 
extended objects, parametrized by coordinates denoted $\sigma$, we use 
indices $\mu$, $\nu$, $\ldots$ for (curved) world-volume indices and $a$, 
$b$, $\ldots$ for tangent space ones. Spacetime curved and flat indices 
are related by the vielbeins $E^{M}_{\ A}$, while for those in the 
world-volume we use $e^{a}_{\ \mu}=\partial_{\mu} X^{M}E_{M}^{a}$. 
Finally $\a,\b,...$ denote spinorial indices}
\beqa
\{Q_\a, Q_\b \}=(\Gamma^M)_{\a\b}P_M+\left. (\Gamma_{M_1\cdots 
M_p})_{\a\b} Z^{M_1\cdots M_p}\right|_\xi
\eeqa 
where $\xi$ denotes the tangent plane to a generic point of the 
$p$-extended object's world-volume and $\Gamma_{M_1\cdots M_p}$ is the 
antisymmetrized product of gamma matrices.

Contracting this expression with a Killing spinor of the background $\e^\a$, 
we reach the well known condition
\beqa
\label{ordinaryBPS1}
(Q\e)^2=K^M P_M + \left. \Phi_{M_1\cdots M_p}Z^{M_1\cdots M_p}\right|_\xi \ge 0
\eeqa
where we have defined 
\beqa
\label{Kdef}
K^M &=& \ov{\e}\,\Gamma^M \,\e \nonumber \\
Z^{M_1\cdots M_p}&=& dX^{M_1}\wedge \cdots \wedge dX^{M_p}.
\eeqa
After this contraction the central extension becomes a $p$-form built up 
as a bispinor\footnote{\label{lafootnote} In certain cases, the $p$-forms 
constructed as bispinors contain extra matrices acting on the underlying 
pair of 10d supergravity spinors. For the sake of generality, we use the 
above generic form. In each particular case, the exact definition of the 
calibration form as a bispinor can be readily obtained from the 
supertranslation algebra.}
\beqa
\label{Phiform}
& \Phi_{(p)}\, =\, \Phi_{M_1\cdots M_p}\, dX^{M_1}\wedge \cdots \wedge 
dX^{M_p} & {\rm with} \ \ \ \  \Phi_{M_1\cdots M_p}\,=\,
\ov{\e}\,\Gamma_{M_1\cdots M_p}\,\e 
\label{firstphi}
\eeqa
General considerations imply that this central extension of the 
supertranslation algebra must be closed, as explicitly follows from the 
Killing spinor equations for $\e$. Its integral over the $p$-dimensional 
spatial brane volume provides the topological charge of the extended 
object. Therefore (\ref{ordinaryBPS1}) is the well-known BPS condition 
for $p$-extended objects. In the rest frame of the object, where 
$P_\mu=(\Hh,0,\cdots,0)$ with $\Hh$ the hamiltonian (energy) density, the 
inequality becomes \footnote{More precisely, $\Hh \ge \left 
|\Phi_\xi\right |$, in order to account for extended objects of opposite 
charges, namely $p$- and anti-$p$ extended objects. For concreteness in 
our examples we provide $\Phi_\xi$ without the absolute value, and
implicitly assume a particular choice of orientation.} \footnote{Note 
that this is a local relation, involving the energy and charge densities. 
Once integrated over the world-volume, it relates the total energy and 
charge.} 
\beqa
\Hh \ge \Phi|_\xi
\eeqa
The condition that the object preserves supersymmetry in the background is 
that the above inequality is saturated.

Now note that for the case of a D$p$-brane in a purely metric background, 
its energy comes just from the pure Nambu-Goto piece of its 
Dirac-Born-Infeld action, namely its energy density is basically
\beqa
\Hh=\sqrt{-\det P[G]}\, \ vol|_\xi
\eeqa
where $P[G]$ represents the pullback of the ambient metric on the 
(spacelike) world-volume directions, and $vol|_{\xi}=d\sigma^1\wedge\cdots 
d\sigma^p$ is the volume form in the world-volume of the $p$-extended 
object. Moreover, the pullback of the form $\Phi$ in (\ref{Phiform}) on 
the tangent plane to the brane can be shown to be
\beqa
\label{curvflat}
&& \left . \Phi_{(p)}\right |_\xi= \sqrt{-\det P[G]}\  
\tilde{\Phi}_{(p)} \nonumber \\
&& {\rm with} \,\, \ (\tilde{\Phi})_{a_1\cdots a_p}=\ov{\e}\,
\Gamma_{a_1 \cdots a_p}\,\e 
\eeqa
Then the above discussion directly leads to calibrations \footnote{An 
alternative viewpoint on calibrations vs. supersymmetry, which we will 
not exploit, is based on kappa symmetry; see, e.g. \cite{kappacali}.}. We 
have a closed form 
$\tilde{\Phi}_{(p)}$ that 
satisfies $\tilde{\Phi}_{(p)}|_\xi \le vol_\xi$ and supersymmetry is 
preserved when the inequality is saturated. Thus supersymmetric branes 
wrap $p$-dimensional submanifolds calibrated with respect to 
$\tilde{\Phi}_{(p)}$. 

Note that a calibrated submanifold is volume minimizing within its 
homology class. Hence it implies the familiar statement that a brane 
carrying no homology charge cannot have non-zero volume and preserve 
supersymmetry. We will see in latter sections that this property does not 
hold in more general backgrounds, with supersymmetric branes described by 
generalized calibrations.

\subsection{Generalized calibrations}
\label{generalgencal}

In the previous section we have seen that constructing $\Phi$ in 
(\ref{firstphi}) with a Killing spinor $\e$ such that $\nabla_\nu \e=0$ 
guarantees that the form is a calibration and, in particular, that it is 
closed. However, in more general supersymmetric supergravity solutions, 
not only the metric but other backgrounds, like the dilaton and NSNS and 
RR field strength fluxes, are turned on. In such backgrounds, the 
covariant derivative of the Killing spinor is not zero, but is related to 
the additional fields. For a general background, the Killing spinor 
equation is obtained by imposing that the supersymmetric variations of 
the fermions vanish. In type IIB theory \footnote{In this section, for 
concreteness we center on type IIB theory, but the conclusions easily 
generalize to type IIA theory.}, following the notation of 
\cite{ortin,Ortinbook}, they are given by
\beqa
\delta_\e \psi_M=\nabla_M \e-\frac{1}{8}H_{M N P}\Gamma^{N P}
\sigma^3 \e+\frac{1}{16}e^{\phi}\sum_{n=1}^{5}\frac{1}{(2n-1)!}
G^{(2n-1)}_{N_1 \cdots N_{2n-1}}\Gamma^{N_1 \cdots N_{2n-1}}\Gamma_M 
\lambda_n\e
\label{susyvar}
\eeqa
with $\phi$ the dilaton and $\lambda_n = \sigma^1 \ ( i\sigma^2)$ for 
$n$ even  (odd). From the above equation one can obtain the expression of 
$\nabla_M \e$ in terms of the background fields.

It is clear that, using the above Killing spinor to construct 
$\Phi_{M_1\cdots M_p}=\ov{\e}\,\Gamma_{M_1 \cdots M_p}\,\e$, as in 
previous section, the resulting form is not closed. On the other hand, 
its exterior derivative will be related to the background fields. This 
points towards a generalization of the concept of calibration, valid for 
general supersymmetric backgrounds. Namely, a \textit{generalized 
calibration} is a form containing a non-closed piece associated to the 
volume form of tangent planes, but involving additional pieces related to 
the additional backgrounds, and completing it to a closed form.

Such a generalized calibration will no longer calibrate minimal volume 
submanifolds, but minimal action world-volumes. A brane wrapping a 
generalized calibrated submanifold minimizes its action (which is no 
longer just the Nambu-Goto action but contains couplings to the 
additional background fields), and preserves some of the supersymmetry of 
the background. Some partial results on these generalized calibrations 
have been obtained in e.g. \cite{gpt,NS5s}. Our aim is to derive and 
describe generalized calibrations more systematically, and including more 
general backgrounds and examples.

A possible strategy to construct generalized calibrations has been proposed in
\cite{DouglasSmith} in certain very supersymmetric backgrounds, based on 
ideas in \cite{GauntPakis}. The point is to consider the non-closed form 
$\ov{\e}\,\Gamma_{M_1 \cdots M_p}\,\e$, and to compute its exterior 
derivative in terms of the additional backgrounds, by using the Killing 
spinor equations. The result then \textit{suggests} how to complete the 
initial non-closed form to a closed generalized calibration.

The strategy we follow is different, and far more systematic. As in 
previous section, we work at the level of the superalgebra, which for a 
general type IIB supergravity background can be written 
\beqa
\{Q_\a, Q_\b \}=(\Gamma^M)_{\a\b} P_M+\sum_{n=1}^{5}(\Gamma_{M_1\cdots 
M_{2n-1}})_{\a\b}
Z^{M_1\cdots M_{2n-1}}_{RR}+(\Gamma_M)_{\a\b} Y^{M}_{NS}+
(\Gamma_{M_1 \cdots M_9})_{\a\b}Y^{M_1 \cdots M_9}_{NS} \quad
\eeqa
This has the familiar structure, with one term involving the generator of 
translations, and central charges (to be made explicit below) related
to possible extended objects in the background.

Once contracting with a Killing spinor $\e$, the right hand side becomes 
a sum of forms of different degrees. Again these forms, denoted 
$\Theta_{RR}$, and $\Pi_{NS}$, characterize the different supersymmetric 
objects in the background. In the rest frame, we obtain the BPS 
inequality
\beqa
\label{geneBPS}
\Hh \ge \sum_{n=1}^{5}\left. \Theta_{(2n-1)}^{RR}\right|_{\xi} +\left. 
\Pi_{(1)}^{NS}\right|_{\xi}+\left.\Pi_{(9)}^{NS}\right|_{\xi}
\eeqa
(where it is understood that in the right hand side only the contributions 
associated to the relevant object are present). The explicit expression 
of the calibrating forms $\Theta$ and $\Pi$ in terms of the background 
fields is given by\footnote{The reader may note that in our conventions 
(see appendix \ref{app}) there is an extra minus sign in the second term 
compared with the corresponding expression in the references. Note 
that in those expressions there is a hidden minus sign in the 
contraction $K^\rho C_{\mu_1 ... \mu_{2n-1}\rho}=-K^\rho C_{\rho\mu_1 ... 
\mu_{2n-1}}$ that is absent in ours, so that the relative signs are the 
same.}  \cite{ortin,Ortinbook}
\beqa
\label{centchar}
\left. \Theta^{RR}_{(2n-1)}\right |_{\xi}&=& i_{K}C_{(2n)} - e^{-\phi}\left 
. \Phi_{(2n-1)}\right |_{\xi}-C_{(2n-2)} \wedge K|_{\xi}\nonumber \\
\left . \Pi_{(1)}^{NS}\right |_\xi&=&i_K B^{NS}_{(2)}+K|_{\xi}\nonumber \\
\left . \Pi_{(9)}^{NS}\right |_\xi&=&i_K B^{NS}_{(10)}+e^{-2\varphi}\left. 
\Phi_{(9)}\right |_{\xi}
\eeqa 
where the 1-form $K$ was defined in (\ref{Kdef}), and $i_{K}C_{p+1}$ is a 
$p$-form defined by $(i_{K}C_{p+1})_{a_1\cdots a_{p}}=
K^{\rho}(C_{p+1})_{\rho a_1\cdots a_p}$. The forms $\Phi_{(p)}$ are 
defined as in (\ref{Phiform}) \footnote{In this particular case, the IIB 
supertranslation algebra dictates that 
$\Phi_{2n-1}=\ov{\e}\lambda_n\Gamma_{\mu_1 \cdots \mu_{2n-1}}\e$ with
$\lambda_n$ defined after (\ref{susyvar}), see footnote 
\ref{lafootnote}.}, and $\Phi|_{\xi}$ denotes its pullback to tangent 
space indices.

An important remark about the above expression is that it is an equality 
between forms on the brane world-volume. Therefore, we will be careful
to translate every piece to world-volume flat indices.

Similarly to the purely metric case, expression (\ref{geneBPS}) allows us 
to interpret the different central extensions of the superalgebra as 
calibrating forms of the world-volume of extended objects. The saturation 
of the BPS inequality is the condition for supersymmetry, and implies that 
the energy minimizing and supersymmetric extended objects are those that 
wrap submanifolds calibrated by one (or a linear combination) of the 
forms $\Theta^{RR}_{(p)}$, $\Pi^{NS}_{(p)}$.

In order to understand the generalized calibration structure of these 
forms, let us center on the concrete case of a D$p$-brane, for which the 
above central extension form reads
\beqa
\label{caligene}
\Theta_{(p)}^{RR}=i_{K}C_{(p+1)}-e^{-\phi}\sqrt{-\det P[G]}\ 
\tilde{\Phi}- C_{(p-1)}\wedge K
\label{gencalibr}
\eeqa
where $(\tilde{\Phi})_{a_1\cdots a_p}=\ov{\e}\, {\cal P}_n\, \Gamma_{a_1 
\cdots a_p}\e$ is the standard calibrating form, associated to the volume 
form. The factor of $\sqrt{-\det P[G]}$ is required for the 
translation of world-volume curved to flat indices (see expression 
(\ref{curvflat})).

As pointed out at the beginning of the section, the closed calibration 
forms that enter in the superalgebra are not just a bilinear of spinors 
with a product of gamma matrices. Rather, as is explicit in the above 
expressions,
there are extra terms that make the form closed. Equivalently, the closed
calibrating form is not related to just the volume of submanifolds, but rather
to the action of D-branes wrapped on them. The branes are supersymmetric when
they saturate the BPS inequality, namely when their world-volume action is
minimized and is given by the integral of the closed calibrating form. 
Therefore, each of the terms appearing in (\ref{centchar}) should 
correspond to a world-volume coupling of the associated extended object. 
Let us see that this is indeed the case for the forms $\Theta_{(p)}^{RR}$
that calibrate D$p$-branes.

The first term in expression (\ref{centchar}) is the ordinary Wess-Zumino 
term $\int_{Dp} C_{p+1}$, adapted to the fact that we are just considering 
the spatial directions of the brane. The second term realizes the 
ordinary volume term appearing in the DBI part of the action. 
Note that interestingly the dilaton prefactor in front of the volume term 
is also obtained.

Concerning the last term in $\Theta^{RR}_{(p)}$ its interpretation as a 
world-volume action term seems to be more involved. In the presence of a 
$C_{(p-1)}$ background, its field-strength acts as a source for the gauge 
field on a D$p$-brane world-volume, $\int_{W_{p+1}} F_{p}\wedge A_1$, 
leading to a world-volume tadpole for the latter. Cancellation of the 
tadpole requires the presence of fundamental strings ending on the 
D$p$-brane, leading to an additional contribution to the tension of the 
system. This is described precisely by the additional contribution to the 
central charge. This issue arises e.g. for D0-brane probes in the 
presence of the background created by D8-brane (or viceversa), and has 
been discussed in detail in \cite{bss} (see \cite{ohta} for similar discussions in dual systems). It will not play any role in our 
examples, so we ignore it in further discussions.

We would like to finish this section with a remark. As we have just 
argued, in a general supergravity background the closed form that 
calibrates supersymmetric brane world-volumes is not only sensitive to 
the volume of the submanifold, but to a number of other terms 
(corresponding to additional couplings of the brane to the background). 
Hence, contrary to the intuition with standard 
calibrations, branes carrying no topological charge (e.g. wrapping   
homologically trivial cycles) may be supersymmetric even if they have 
non-zero volume. This beautiful feature of having stable and supersymmetric 
D$p$-branes carrying no charges \footnote{Such branes have appeared in 
different contexts, for instance \cite{gravitons,kr,kk,denef}.} leads to 
novel phenomena and allows for new model-building possibilities. We will 
explore some of them in the following sections. 

\subsection{Branes within branes}
\label{gensitu}

There are brane systems with more than one contribution to central charges. 
These include simple superpositions of different branes, preserving some common
supersymmetry, but also bound states of branes. A familiar class of 
examples of the latter are obtained for D-branes with topologically 
non-trivial world-volume gauge bundles, or in the presence of 
NSNS $B$-fields. As follows from the above general discussion, the 
generalized calibration characterizing the supersymmetry properties of 
these bound states is obtained as a superposition of the corresponding 
central charges.

However, it will be useful for later sections to describe a more compact 
expression for these superpositions of central charges, for systems of 
D-branes with lower-dimensional induced charges. We consider the 
generalized calibration associated to a D$p$-brane with a world-volume 
gauge field strength\footnote{Calibrations for bound states  of branes in the particular context of M-theory (and fivebranes in IIA) can be found in e.g. last reference of \cite{martelli}.} $F$, and in the presence of a NSNS 2-form field $B$. 
Following the general discussion above, the generalized calibration 
should contain a piece describing the tension of the bound state, and a 
piece describing the coupling $i_K C_{(n)}$ for each of the charges 
present.

These two pieces can be immediately read from the structure of the 
general D-brane action with the above described backgrounds. Hence, the 
generalized calibration encoding the information of such bound states is 
given by
\beqa
\label{Bgencali}
\left . \Theta_{(p)}^{RR}\right |_{\xi}= i_{K}\left . \left ( 
\sum_{n}C_{(n+1)}\wedge e^{(B+F)}\right )\right|_\xi - e^{-\phi}\left . 
\sqrt{-\det \left (P[g+B]+F \right )}\tilde{\Phi}_{(p)}\right|_{\xi}
\eeqa
where 
$\tilde{\Phi}$ is defined after (\ref{caligene}). 

Clearly the first term describes a formal sum of the central charges 
associated to the coupling of lower-dimensional branes to the RR gauge 
potentials, and the second describes the tension of the bound state. This 
expression was already implicit in the description of the supersymmetry 
properties of branes in \cite{marino} from the $\kappa$-symmetry viewpoint.

The usefulness of the above expression will become more manifest in 
sections \ref{warmup} and \ref{general}, where we consider D-branes in 
NSNS (and RR) flux backgrounds. However, it is important to notice a 
general drawback of the above expression: since it exploits induced 
charges and tensions on the D-brane volume, it over-emphasizes the role of 
the D-brane, whereas a true calibration is expected to be a form defined 
over spacetimes, which is ultimately pullbacked onto D-brane volumes. The 
subtlety is not relevant for situations with constant $B$-fields, but some 
effects may arise if non-trivial NSNS 3-form flux. Indeed, in some 
instances (see section \ref{fixing}) the above expression rather provides 
the actual action of a D-brane configuration, which only when minimized 
provides the value of the generalized calibration over the D-brane. 
For configurations attaining this minimum, the D-branes are generalized 
calibrated.

\section{Some examples of generalized calibrated branes}
\label{examples}

\subsection{Supergravity backgrounds}

In this section we apply the above general discussion to some examples. 
A good testing ground for these ideas is provided by the supergravity 
backgrounds created by stacks of branes. For instance, consider the 
supergravity solution for a stack of $N$ D$p$-branes \footnote{As usual, 
this expression is valid for $p\le 6$. For $p>6$ the expression of the 
harmonic form changes, and the solution provides a local description of 
the spacetime, which is no longer asymptotically flat. However, our 
discussion can be applied to the latter cases as well.} (similar ideas 
may be discussed for other NS- or M-branes, or other supersymmetric 
backgrounds), given by \cite{polchinski2}
\beqa
ds^2 & = & H(x)^{-1/2} \eta_{\mu\nu} dx^\mu dx^\nu +
H(x)^{1/2} dx^m dx^m \nonumber \\
e^{2\phi} & = & H(x)^{(3-p)/2} \nonumber \\
C_{p+1} & = & (\,H(x)^{-1}-1\,) \, dx^0 \ldots dx^p
\label{Dpbrsol}
\eeqa
Here $H(x)$ is a harmonic function in the transverse space. For a 
point-like source $H=1+g_sN(\a')^{(7-p)/2}/r^{7-p}$, with $r^2=\sum_m (x^m)^2$,
but more generally some of such backgrounds may correspond/be mimicked by 
other kinds of sources, like fluxes, or distributions of branes and fluxes, 
by simply considering more general harmonic functions $H(x^m)$.

We are interested in considering brane probes in these backgrounds. We 
may center on D$p'$-brane probes for concreteness (with analogous results 
for NS-branes). The generalized calibration form associated to a probe 
D$p'$-brane is given by (\ref{gencalibr})
\beqa
\Theta_{(p')}^{RR}=i_{K}C_{(p'+1)}-e^{-\phi}\sqrt{-\det 
P[G]}\;\,\tilde{\Phi}_{(p')}
\eeqa
where we have dropped the last term which, as anticipated, plays no 
r\^ole in the situations we are to consider. Recall that $P[G]$ 
represents the pullback of the ambient metric on the $p'$  
\textit{spacelike} components of the world-volume. Also $K$, 
$\tilde{\Phi}_{(p')}$ are the 1- and $p'$-form defined as spinor 
bilinears with the Killing spinor of the supergravity background $\e$. 

The Killing spinor preserved by a D$p'$-brane background is of the 
form \cite{Ortinbook} 
\beqa
\label{killspin}
\e=H^{-1/8}(r)\e_0
\eeqa
with $\e_0$ a constant spinor normalized to one, $\ov{\e}_0\Gamma^0 
\;\e_0=\e_0^\dagger\e_0=1$, and satisfying $\Gamma_{01\cdots p'}\e_0=\e_0$.

With this expression for the spinor, the second term of the calibration 
becomes 
\beqa
e^{-\phi}\, \sqrt{-\det P[G]}\, H^{-1/4}\ \tilde{\Phi}_0 
\quad \quad {\rm with} \,\, 
\left ( \tilde{\Phi}_0\right )_{a_1\cdots a_{p'}}=\ov{\e}_0\Gamma_{a_1 
\cdots a_{p'}}\e_0,
\eeqa 
Notice that the prefactor is responsible for this volume-related piece 
not being closed. 

From expression (\ref{Dpbrsol}), we have $H^{-1/4}=\sqrt{-g_{00}}$. Hence
it can be absorbed in the square root of the determinant, to give the 
determinant of the induced metric along \textit{all} $p'+1$ world-volume 
dimensions.

Concerning the first term, if we recast
\beqa
K^{M}\,=\,\ov{\e}\,\Gamma^{M}\,\e\,=\,H^{-1/4}(r)\,e^{M}_{\ \ 
a}\,\ov{\e}_0\,\Gamma^{a}\,\e_0
\eeqa
then the non-constant piece is factored out.
The contraction $i_{K}C_{(p'+1)}$ then reads 
\beqa
\left ( i_{K}C_{(p'+1)}\right )_{a_1 \cdots a_{p'}}\, =\, 
H^{-1/4}(r)\,e^{0}_{\ \
a}\,\ov{\e}_0\,\Gamma^{a}\,\e_0 \, C_{0a_1 \ldots a_{p'}}\, = 
\, \ov{\e}_0\,\Gamma^0\,\e_0 \, C_{0a_1 \ldots a_{p'}}\,
\eeqa
In total, we end up with the following general calibrating form for the
D$p'$-brane
\beqa
\Theta_{(p')}^{RR}\, =\,C_{(p'+1)}\ov{\e}_0\,
\Gamma^0\,\e_0\,-\,e^{-\phi}\,\sqrt{-\det P[G]}\ \tilde{\Phi}_{(0)}
\label{gencalibdp}
\eeqa
where now $P[G]$ represents the pullback of the ambient metric on 
the complete $(p'+1)$-dimensional world-volume, and all $\Gamma$'s have 
tangent space indices (and are thus constant). Note that the 
bispinorial factor in the first term is just the normalization condition 
for the spinor $\e_0$, that we have set to 1.

We would like to point out that the manipulations carried out to reach the 
final form of the generalized calibration can be similarly repeated for 
other brane-like supergravity solutions. Hence, examples like those is next 
section can be similarly constructed for generalized calibrations in 
supergravity backgrounds created by NS5-branes, M-branes, etc.

\subsection{Examples}

We may consider several examples of calibrated and generalized calibrated 
submanifolds in these backgrounds, corresponding to supersymmetric branes.

\subsubsection{An example with non-trivial metric and flux}
\label{exmetricflux}

Let us consider a simple example, which is a version of the generalized 
calibrations considered in \cite{gpt}, where both the volume of the cycle 
as well as a RR potential are involved in the calibration. Consider the above
background for $p=3$ 
\beqa
ds^2 & = & H(x)^{-1/2} \eta_{\mu\nu} dx^\mu dx^\nu +
H(x)^{1/2} dx^m dx^m \nonumber \\
C_4 & = &  (\, H(x)^{-1}-1\,)\, dx^0 \wedge \ldots \wedge dx^4 
\label{D3brsol}
\eeqa
with constant dilaton. The particular case of $H(r)=\alpha'^2 g_sN/r^4$ 
corresponds to the maximally supersymmetric background AdS$_5\times S^5$.

Consider a D3-brane spanning the direction 0123, and sitting at a value 
of the coordinates $x^m$. This submanifold in general does not minimize 
the volume, obtained from the familiar volume form induced from the metric
\beqa
\sqrt{-\det P[G]} \,dx^0\, dx^1\, dx^2\, dx^3 \, = 
H(r)^{-1} dx^0\, dx^1\, dx^2\, dx^3 
\label{volformD3}
\eeqa
unless $H(x)=1$, i.e. a flat background. However, the D3-brane {\em is} 
supersymmetric in this background, since it is generalized calibrated 
with respect to the generalized calibration (\ref{gencalibdp}), 
corresponding to the background with flux, which takes the form
\beqa
\Theta_{D3}= C_4 -\,\sqrt{-\det P[G]} \,dx^0\, dx^1\, dx^2\, dx^3 \, =\,
C_4\,-\,H(x)^{-1} \,dx^0\, dx^1\, dx^2\, dx^3\,= \,- dx^0\, 
dx^1\, dx^2\, dx^3 \nonumber \\ 
\eeqa
Notice that the generalized calibration completes the expression 
(\ref{volformD3}) with additional pieces, to yield a closed form.
The D3-brane is supersymmetric since its action is given not just by its 
volume, but rather includes an `electrostatic' energy due to its coupling 
to the background RR 4-form. 

\subsubsection{An exotic standard calibration}
\label{exotic}

Consider the same background (\ref{D3brsol}), now probed by a D5-brane 
spanning the directions 012456. Centering e.g. in the AdS$_5\times 
S^5$ situation, written as
\beqa
ds^2 & = & \left( \frac{r}{R} \right)^2 \eta_{\mu\nu} dx^\mu dx^\nu +
R^2 \, \frac{dr^2}{r^2} \, +\,  R^2\, d\Omega_5^{\, 2} = \nonumber \\
& = & \left( \frac{r}{R} \right)^2 \eta_{\mu\nu} dx^\mu dx^\nu +
R^2 \, \frac{dr^2}{r^2} \, +\,  R^2\, ( \cos^2\theta \, d\Omega_2^{\, 2}+ 
\sin^2 \theta d\Omega_2'^{\, 2})
\eeqa
the D5-brane is spanning an AdS$_4\times S^2$ subspace. Although the D5-brane 
is wrapped on a trivial cycle in the internal manifold, it is 
supersymmetric. Indeed, applying the formula (\ref{gencalibdp}) to this 
situation (and noticing the RR field does not give any contribution), we see that the brane is calibrated by the form 
\beqa
\Theta_{D5}\, =\, -\sqrt{-\det P[G]}\, dx^0\, dx^1\, dx^2\, \, dr\, 
d{\rm 
vol}_{S^2} \, =\, -r^2\, dx^0\, dx^1\, dx^2\, dr\, d{\rm vol}_{S^2} \, =\, 
-dx^0 \, dx^1\, dx^2\, dx^4\, dx^5\, dx^6 \nonumber \\ 
\eeqa
Notice that the form is closed, and indeed corresponds to the volume form 
in flat space. In this sense, the complete submanifold spanned by the 
D5-brane minimizes its volume. On the other hand, regarding the 
configuration as a compactification to 5d, the D5-brane may seem able to 
decrease its volume (and hence its energy) by simply slipping off the 
$S^2$ in the $S^5$. The paradox is solved as follows. As discussed in 
\cite{kr}, the scalar parameterizing the slipping is tachyonic in the AdS$_4$ 
non-compact directions of the D5-brane, but not tachyonic enough to 
violate the Breitenlohner-Friedmann bound \cite{bf}, and hence does not 
lead to an instability. Hence, there is a key role played by the non-compact 
directions in the stability of the system. In other words, the process of 
slipping off the $\IS^2$ requires exciting the spacetime profile of the 
tachyonic field, in such a way that the combined motion increases the 
energy. In this sense, the above configuration is volume minimizing, in 
agreement with it being calibrated. The result of having a stable brane 
wrapped on a trivial cycle is nevertheless surprising and interesting. It 
will play an important role in later sections. Indeed the above calibrated 
submanifolds will be exploited in section \ref{building}

\subsubsection{Generalized hermitian calibrations}
\label{hermitian}

The above two kinds of calibrations (pure metric, and metric plus flux 
calibrations) may be combined. Consider again the background 
(\ref{D3brsol}) and consider D3-brane probes wrapped on 
2-cycles in the directions 2345. The volume form for such 2-cycles is 
\beqa
H(x^m)^{-1}\, dx^0\, dx^1\, dx^2\, dx^3 \, +\, dx^0\, dx^1\, dx^4\, dx^5
\label{hermitfake}
\eeqa
In the $AdS_5\times S^5$ case, the brane wraps an $AdS_3$ slice of the 
whole $AdS_5$ and a circle in the internal space. The second piece in 
the volume form precisely corresponds to the part of the D3-brane spanning 
the $AdS_3\times S^1$ along 0145. 

The total form is not closed, due to the first piece. On the other hand, 
supersymmetric branes must have world-volumes calibrated not with respect 
to (\ref{hermitfake}), but with respect to
\beqa
\Theta_{D3} & = & -H(x)^{-1}\, dx^0\, dx^1\, dx^2\, dx^3 \, -\, dx^0\, dx^1\, 
dx^4\, dx^5\,
+\, C_4\, \nonumber\\ &=& -dx^0\, dx^1\, (\, dx^2\, dx^3 \, +\, dx^4\, dx^5\, ) =2i
\, dx^0\, dx^1\, (\, dz_1\, d{\ov z}_1 \, +\, dz_2\, d{\ov z}_2 \, )
\eeqa
 where we have introduced complex coordinates $z_1=x^2+i x^3$, 
$z_2=x^4+ix^5$. The generalized calibration contains additional pieces, 
providing a closed form.

The last expression makes it manifest that any D3-brane 
spanning 01 and a holomorphic 2-cycle in $z_1$, $z_2$ is generalized 
calibrated. These hermitian (as generalization of holomorphic)  
generalized calibrated submanifolds have been studied in \cite{gpt}.
The extension to other generalized calibrations (generalizing calibrated 
cycles in flat space or other special holonomy manifolds) can be carried 
out analogously \cite{gpt}, so we skip their discussion.

\subsubsection{An example with non-trivial metric and dilaton}

It is interesting to describe generalized calibrated submanifolds in 
supersymmetric backgrounds with varying dilaton. Let us consider one such 
example. Consider the background (\ref{Dpbrsol}) for $p=5$, and a D3-brane 
probe along 0126. This corresponds to a supersymmetric probe, since it 
is generalized calibrated with respect to the calibration
\beqa
\Theta_{D3} \, = \,- e^{-\phi} \, \sqrt{-\det P[G]} \, dx^0\, dx^1\, 
dx^2\, dx^6\, =\, -H(x)^{1/2} H(x)^{-1/2} \, dx^0\, dx^1\,
dx^2\, dx^6\, = \, -dx^0\, dx^1\, dx^2\, dx^6\nonumber \\
\eeqa
The dilaton radial factor is canceled against the metric piece, 
determined by the number of  directions with $H(r)^{\pm 1/2}$ factors. 
Clearly, many other examples are possible. For instance, one can easily 
derive that D3-branes spanning 01 and a 2-cycle in 2367 is generalized 
calibrated if the 2-cycle is holomorphic in the complex coordinates 
$z_1=x^2+ix^3$, $z_2=x^6+ix^7$. 

\subsubsection{An example with non-trivial metric, flux, and dilaton}

Finally, let us describe a simple example where metric, dilaton, and 
RR fields enter in the generalized calibration. Consider the background 
(\ref{Dpbrsol}) for $p=4$, and consider a D4-brane probe along 01234.  
The D4-brane is supersymmetric, since it spans a generalized calibrated 
submanifold with respect to the form
\beqa
\Theta_{D4} & = & C_5\, - e^{-\phi} \, \sqrt{-\det P[G]} \, dx^0\, dx^1\, 
dx^2\, dx^3\, dx^4\,  = \, -dx^0\, dx^1\, dx^2\, dx^3\, dx^4\, 
\eeqa
A similar calibration will appear in section \ref{thedfour}.

\section{Generalized calibrations in flux compactifications: A warm-up 
example}
\label{warmup}

In this section we provide a simple example that illustrates that 
supersymmetric branes in flux compactifications\footnote{It is worth noting that we center on supersymmetric fluxes of the B(ecker)-type (e.g. in \cite{granapolchinski}). It would be interesting to extend this work on generalized calibrations to other classes of supersymmetric fluxes like those appearing in \cite{otherflux}} and compactifications on 
non-Calabi-Yau spaces (but with $SU(3)$ structure) are generalized 
calibrated. A more general description will be provided in next section, 
but it is useful to consider a warm-up example first.

\subsection{The type IIB D3-brane in a flux background}

We are interested in considering type IIB string theory compactifications 
on Calabi-Yau threefolds, with NSNS and RR 3-form field strength fluxes, 
$\Hh_3$ and $F_3$, see \cite{drs,gkp} for the basic description of the 
setup. For simplicity we 
may take a non-compact version,
with topology $\IR^5\times \IS^1$, with $\IR^5$ parametrized by coordinates
$x^\alpha$, and $\IS^1$ parametrized by a periodic coordinate $x$ (on 
which we will eventually apply T-duality). We denote these six 
coordinates by $x^m$. Some additional directions beyond $x$ could be 
considered to be compact, by imposing periodic identifications on the 
corresponding $x^\alpha$'s, as we will consider in certain instances 
(notice that if we consider all directions to be compact, thus describing 
a $T^6$, we will implicitly assume the introduction of orientifold 
O3-planes required for RR tadpole cancellation). Alternatively, the 
configuration can be regarded as a local model of more general situations. 

Let us introduce RR and NS-NS 3-form fluxes $F_3$, $\Hh_3$ such that 
$G_3=F_3-\tau \Hh_3$ (with $\tau$ the IIB complex coupling) is imaginary 
self-dual (ISD)\footnote{Note that in examples with additional compact 
dimensions, the fluxes along compact 3-cycles must be properly 
quantized.}. The fluxes backreact on the background by inducing a warp
factor for the metric and a non-vanishing RR 5-form field strength $F_5$. The
supergravity solution for the configuration is of the black 3-brane form, 
due to the imaginary self-duality of the flux \cite{drs,gkp}.
We consider a flux distribution independent of $x$, so that the harmonic 
function $H$ controlling the solution depends only on the $x^\alpha$. 
The type IIB background is
\beqa
\label{backg1}
ds_{IIB}^2&=& H^{-\oh}(x^\alpha)\, ds^4_{4d}\, +H^{\oh}(x^\alpha)\, dx^m dx^m 
\nonumber \\
C_4&=&(H^{-1}(x^\alpha)-1)\, dvol_{0123} \\
C_2&=&(C_2)_{\a\b}\, dx^\a\, dx^\b\, +\, (C_2)_{x\a}\, dx\, dx^\a  
\Longrightarrow 
F_3\, =\, (F_3)_{\a\b\g}\, dx^\a\, dx^\b\, dx^\g\,+\,(F_3)_{x\a\b}\, dx\, 
dx^\a\,dx^\b
\nonumber \\
\Bb_2&=&(\Bb_2)_{\a\b}\, dx^\a\, dx^\b\, +\, (\Bb_2)_{x\a} \, dx\, dx^\a 
\Longrightarrow 
\Hh_3\, =\, (\Hh_3)_{\a\b\g}\, dx^\a\, dx^\b\, dx^\g\, +\, (\Hh_3)_{x\a\b} 
\, dx\, dx^\a\,dx^\b\nonumber 
\eeqa
where $dvol_{0123}=dx^{0}\wedge dx^1\wedge dx^2\wedge  dx^{3}$, and a 
convenient gauge has been chosen for $\Bb_2$, $C_2$. In principle we leave 
the explicit form of $H(x^\alpha)$ arbitrary, so that the configuration 
describes a general class of situations (including distant sources, like 
additional O3-planes or D3-branes).

Under some additional conditions (namely, that the flux $G_3$ is (2,1) and 
primitive), the above background is supersymmetric. Now consider we place 
on it a probe D3-brane along 0123. This preserves the same supersymmetry 
as the background, and indeed it corresponds to a brane on a generalized
calibrated submanifold of 10d spacetime. The discussion is similar to that of 
a D3-brane in the presence of a D3-brane supergravity background, discussed in
section \ref{exmetricflux}. The D3-brane is generalized calibrated, with 
respect to the form
\beqa
\Theta_{D3} \, =\, C_4\, - \, H^{-1}(x^\alpha)\, dvol_{0123}
\eeqa
The generalized calibrated nature of the D3-brane simply reflects the
supersymmetric  cancellation between the gravitational and RR 4-form
interactions of the D3-brane with the background created by the fluxes.

The fact that such D3-branes are generalized calibrated is perhaps not 
surprising, since the class of the point is a homologically non-trivial 
0-cycle in the Calabi-Yau, and the generalized calibration 
seemingly describes a mildly deformed version of the corresponding 
standard calibration on the underlying Calabi-Yau \footnote{In fact, 
neglecting the flux backreaction, which is mimicked by taking constant 
$H$, the generalized calibration becomes a standard calibration.}. 
However, the fact that the D3-brane is generalized calibrated is really 
crucial, since D3-brane charge in the presence of fluxes is a torsion 
class in K-theory, and hence such D3-branes do not carry  $\IZ$-valued 
charges (or even no charge at all, so that they are topologically 
trivial). Thus the only way in which the latter systems can be 
supersymmetric is to correspond to {\em generalized} calibrations. We 
will return to this point in section \ref{ktheory}.

\subsection{Generalized calibrations in non-CY backgrounds from T-duality}
\label{thedfour}

T-duality acts in a non-trivial way on backgrounds with 3-form fluxes, by 
transforming certain components of the NSNS flux into curvature contributions 
in the T-dual background. Hence, we can exploit T-duality to explore 
supersymmetric brane wrappings in these (non-Calabi-Yau) geometries.

Let us T-dualize the above background together with the probe D3-brane 
along the isometric internal direction $x$. In the discussion below, the 
IIA and IIB RR potential and field strengths are distinguished by their 
degree, while for the NSNS ones, we denote the IIB quantities by 
$(\Bb_2,\Hh_3)$ and the IIA quantities by $(B_2,H_3)$. We also introduce 
the definitions in (\ref{defiKSTT}).

Using standard T-duality formulae (see appendix), the resulting 
T-dual IIA metric 
is
\beqa
ds_{IIA}^2\, =\, H^{-\oh}\, \left(\,  ds^2_{0123}\, +\, 
(dx+g_{(x)}\,)^2\right)\,+\,H^{\oh}\, dx^\a dx^\a
\label{tdualmetric}
\eeqa
with
\beqa
\label{g(x)w}
g_{(x)}\, =\,\frac{g^{IIA}_{x\a}}{g^{IIA}_{xx}}\,dx^\a\,=\,-\Bb_{x\a}\,dx^\a 
\quad \quad ; \quad \quad
\om \,=\,-dg_{(x)}\,=\,-(\Hh_3)_{(x)\a\b}\, dx^\alpha dx^\beta
\eeqa
The NSNS 2-form and dilaton are
\beqa
B_2&=&(\Bb_2)_{\a\b} \, dx^\a dx^\b \nonumber \\
e^{\phi^{IIA}}&=&\frac{e^{\phi^{IIB}}}{\sqrt{g_{xx}^{IIB}}}=H^{-1/4}
\eeqa
Concerning the RR potentials, we do not need to compute all of them in 
the dual side. As will become clear in what follows, the dual object to 
our original D3 is only sensible to the RR 5-form potential with 
components $(C_5)_{0123x}$, which turns out to be
\beqa
(C_5)_{0123x}&=&[(C_4)_{0123}]_x=(H^{-1}-1)\, dvol_{0123x}
\eeqa
The resulting metric (\ref{tdualmetric}) has the same structure as that of 
a black 4-brane, but with a non-trivial twisting given by $g_{(x)}$. In 
the compact case, the compact manifold is no longer a $T^6$ but a twisted 
torus. In particular, the $x$-coordinate is non-trivially fibered over 
all those two tori parametrized by $(x^\a,x^\b)$ such that $(\Hh_3)_{\a\b 
x}\ne 0$. For each pair $(x^\a,x^\b)$ the first Chern class of the 
non-trivial fibration is precisely given by the (quantized) integral of 
$\Hh_3$ in the $T^3$ spanned by $x$, $x^\a$, $x^\b$. This non-trivial 
fibration makes the topology of the resulting space quite peculiar, see, 
e.g. \cite{kstt,glmw}. For our purposes it is enough to mention that the 
dual manifold is neither K\"ahler nor complex (hence non Calabi-Yau) but 
nevertheless leads to supersymmetric flat Minkowski spacetime, as 
expected from the dual side.

Concerning the probe D3, its T-dual corresponds to a D4-brane spanning 
the 4d Minkowski directions, and wrapped on the circle fiber parametrized by
$x$. The twisting on the metric implies that, in situations with compact
dimensions, this 1-cycle corresponds to a torsion class. Consider a 
3-cycle given by the circle fibration over a 2d space $\Sigma$ on the 
base, such that $\int_\Sigma dg_{(x)}=K\in \IZ$. Then the twisting 
implies that $K$ times the class of the circle 1-cycle is 
trivial in homology, and hence $K$ wrapped D4-branes can unwind and disappear 
\footnote{\label{slipping} The argument for this is very similar to the 
unwinding of the $\IS^1$ fiber in the Hopf fibration of $\IS^3$ over $\IS^2$
(this is similar to the case $K=1$, for other values of $K$, one may just take
the Lens space $\IS^3/\IZ_K$). Namely, one can start with a D4-brane wrapped 
$K$-times over the $\IS^1$, and sitting at a point on the base 2-cycle 
$\Sigma$. Next, deform it so that its projection on the base is a small 
circle, and then grow the latter until is has swept out the whole of 
$\Sigma$. By then the projection is a small circle `on the opposite side' 
of $\Sigma$, and the twisting has managed to unwind the cycle in the $x$ 
direction, so that it can safely shrink to zero size and disappear.}. We 
will come back to the unwinding process in section \ref{ktheory}.

The fact that the 1-cycle parametrized by $x$ is a torsion class in 
homology implies that D4-branes wrapped on them do not carry any 
$\IZ$-valued charge. On these grounds, one would be tempted to propose 
that they are not BPS, and that $K$ of these D4-branes dynamically decay 
to the vacuum. On the other hand, the T-duality with D3-branes suggests 
the D4-branes should be supersymmetric, and hence BPS and stable. The 
resolution of this seeming paradox lies in generalized calibrations. 
Indeed, the D4-branes wrap a generalized calibrated submanifold, and 
hence are BPS and stable, even if they do not carry any charge. 

Following our general discussion, the generalized calibration for these 
D4-branes is
\beqa
\Theta_{D4} =\,C_5\, - \,e^{-\phi} \sqrt{-\det P[G]}\, d\,vol_{0123}\wedge dx 
\eeqa
which upon straightforward computation reads
\beqa
\left . \Theta_{D4}^{RR}\right |_{D4}\, =\, -dx^0\, dx^1\, dx^2\, dx^3\, dx
\eeqa
showing that the D4-branes are indeed generalized calibrated.

\subsection{Decay of D4-branes in trivial cycles, and T-dual of K-theory}
\label{ktheory}

We cannot refrain from making a small detour and discussing the process of
unwinding of objects wrapped on the $\IS^1$ fibers, like the above D4-branes. 
As discussed in footnote \ref{slipping}, a D4-brane wrapped over the 
$\IS^1$, and sitting at a point on the base 2-cycle $\Sigma$, can be 
continuously deformed on the base until is has swept out the whole of the 
two-dimensional submanifold $\Sigma$, so that the winding is finally 
undone by the twisting of the fibration of $\IS^1$ over $\Sigma$. This 
process violates D4-brane number in $K$ units, where $K$ is the first 
Chern class of the $\IS^1$ fibration over $\Sigma$.

This process can also be regarded as mediated by a brane instanton 
(equivalently, there exist domain walls separating the two 
configurations, with and without D4-branes). The branes mediating the 
process are in fact D4-branes spanning the two-dimensional base $\Sigma$, 
times a codimension one 3-plane in 4d Minkowski space (so that they are 
to be interpreted as instantons or domain walls, depending on whether 
they are transverse to the time or space coordinates). 

This description raises a number of interesting points. First, because of 
the twisting of the $\IS^1$ fiber transverse to the instanton D4-brane, there 
is a world-volume topological charge of $K$ units. The world-volume 
tadpole is canceled if there are exactly $K$ 
D4-brane wrapped on the $\IS^1$ fiber, and `ending' on the instanton 
D4-brane \footnote{More precisely, the wrapped and the instanton 
D4-branes join smoothly, in a geometry locally reproducing the smooth 
intersection $xy=\epsilon$.} on $\Sigma$. This is exactly the feature 
required for the instanton to violate D4-brane winding number in $K$ units.

A second interesting point is the following. In compactifications with 
non-CY geometries of the above kind, there is a RR tadpole contribution 
from the flux (T-dual of the familiar one of type IIB 3-form fluxes).
Namely, due to the twisting of the geometry, there is a 10d coupling (see 
section 7.1 of \cite{cascur})
\beqa
\int_{10d} dg_{(x)} \wedge F_{4(x)} \wedge C_5
\eeqa
which leads to a tadpole for $C_5$, canceled against local sources.
The instanton D4-brane we are studying is a source of RR 5-form $C_5$, 
namely the flux of $F_4$ through the 4-cycle dual to $\Sigma$ jumps by one 
unit. This implies that in crossing the instanton, there is an 
increase in $K$ units to the $C_5$-tadpole, arising from the interplay of 
the additional unit of $F_4$ with the $K$ units of $dg_{(x)}$. This 
implies that the disappearance of the $K$ D4-branes is accompanied by an 
 increase of the flux contribution to the tadpole, so that tadpole 
cancellation is maintained. 

\medskip

It is interesting to relate the above process to the T-dual type IIB 
picture. Mapping objects and couplings in a simple way, the type IIB 
version of the process corresponds to the disappearance of $K$ D3-branes 
via a brane instanton corresponding to a D5-brane wrapped on a 3-cycle 
with $K$ units of $\Hh_3$ flux. This has been described in \cite{kpv}, 
and in \cite{klst} in the domain wall picture. The process also involves 
an increase of $F_3$-flux through the dual 3-cycle, so that the 
$C_4$-tadpole increase by $K$ units, compensating the disappearance of 
D3-branes, and ensuring tadpole cancellation after the decay. In fact 
this process is the brane instanton process introduced in \cite{mms} to 
explain the decay of branes whose charge is non-trivial in 
homology, but trivial in K-theory. It is extremely amusing to see that in 
the type IIA language the whole discussion can be carried out purely in terms 
of homology, thus making it more geometrical and intuitive.

\subsection{Generalized calibrated branes and mirror symmetry}
\label{sectD6(1)}

In this section we intend to generalize the arguments in the previous 
sections to the case of 3 T-dualities, thus mirror symmetry \cite{syz}, 
for a particular class of backgrounds, studied in \cite{glmw}. We 
consider certain type IIB theory in flux backgrounds and carry out three 
T-dualities, such that each $\Hh_3$ component has at most one leg along 
the T-duality directions. The resulting geometries are half flat manifolds 
\cite{glmw}, i.e. a particular relatively simple class of manifolds with 
$SU(3)$-structure. Hence, they serve as a good testing ground to explore 
the possibilities of brane wrapping, an issue in which we show 
generalized calibrated subspaces are crucial.

Our aim is to follow the same strategy as above to find an example of 
those generalized calibrated branes. Namely, T-dualize a D3-brane in a 
supersymmetric flux background, along three isometric directions. The 
mirror picture is a D6-brane wrapping a generalized calibrated 
submanifold of a half-flat geometry\footnote{An example of D6-brane wrapping a 3-cycle on a (non half-flat) SU(3)-structure manifold is discussed in \cite{tatar}}.

For this case, we slightly change the notation with respect to previous 
sections. We consider at least three compact dimensions, denoted $x^\a$, 
along which we will T-dualize the configuration, and three additional 
ones, which may be non-compact, denoted $y^\a$.

For concreteness, let us consider a type IIB background with ISD 
supersymmetric 3-form fluxes of the form
\beqa
\Hh_3&=&(\Hh_3)_{y^\a y^\b x^\g} \, dy^\a dy^\b dx^\g \nonumber \\
F_3&=&(F_3)_{x^\a x^\b y^\g}\, dx^\a dx^\b dy^\g 
\eeqa
This is a restricted class of fluxes, which is ISD for $\tau=i$. 
Moreover, there
are supersymmetric examples within it, e.g.
\beqa
\Hh_3&=& -dx^1 dy^1 dy^3\, +\, dx^2 dy^2 dy^3 \nonumber \\
F_3&=& dx^1 dy^1 dx^3 - dx^2 dy^2 dx^3
\eeqa
which in fact preserves $N=2$ supersymmetry.

Since we take the coordinates $\vec{x}$ to be isometric, the most 
appropriate gauge to chose is:
\beqa
\Bb_2 & = & (\Hh_3)_{y^\a y^\b x^\g}\, y^\a\, dy^\b dx^\g\nonumber \\
C_2 & = & (F_3)_{x^\a x^\b y^\g}\,y^\g\, dx^\a dx^\b
\eeqa
Again the presence of the fluxes generates a black 3-brane background of 
the form
\beqa
\label{metricD6(1)}
ds^2&=&H^{-\oh}\, ds^2_{4d}\,+\,H^{\oh}\,\left( \, d\vec{x}^2 \,+\, d\vec{y}^2 
\, \right ) \nonumber \\
C_4&=&(H^{-1}-1)\, dvol_{0123}
\eeqa
where we take the harmonic function $H=H(\vec{y})$ to be independent of 
the $x^\a$. By T-dualizing along the three coordinates $x^\a$ as in the 
previous section, the resulting metric and dilaton in the IIA side read
\beqa 
ds^2 & = & H^{-\oh}\, \left(\, ds^2_{4d}\, +\, \sum_{\a=1}^{3} \,(\,dx^\a 
+g_{(x^\a)}\,)^2\, \right) \,+\, H^{\oh}\, d\vec{y}^2 \nonumber  \\
e^{\phi}&=&H^{-3/4}
\label{halfflatmetric}
\eeqa
where we defined $g_{(x^\a)}$ in terms of the original fields as in 
(\ref{g(x)w}), namely $g_{(x^\a)}=-\Bb_{x^\a y^\b}dy^{\b}$. In the dual 
side, for the particular choice of $\Hh_3$ we have considered, no 
$B$-field component survives.

Following \cite{glmw}, this realization of mirror symmetry should yield a 
half-flat manifold in the T-dual. This can be verified by defining the 
tangent complex 1-forms
\beqa
e^\a &=&H^{-1/4}\left ( dx^\a+g_{(x^\a)} \right ) + i H^{1/4}\, dy^\a 
\label{frames}
\eeqa
and constructing the 2- and 3-forms 
\beqa
J \, =\,- \frac{i}{2} \left ( e_1\wedge \ov{e_1} + e_2\wedge \ov{e_ 2} + 
e_3\wedge \ov{e_ 3}\right ) \quad ;\quad
\Omega  =  e_1\wedge e_2\wedge e_3
\label{jomega}
\eeqa
These are globally defined but not closed. One easily checks that they 
satisfy the conditions
\beqa
d(J\wedge J)=0 \quad \quad ; \quad \quad d(\pim \Omega)=0
\eeqa
which are the conditions defining a half-flat manifold.

Concerning the dualization of the RR gauge potentials, we need not carry 
it out in full generality for our purpose. The only component relevant 
for our discussion below is
\beqa
(C_7)_{0123x^1x^2x^3}=(H^{-1}-1)\, dvol_{0123} \wedge dx^1 \wedge dx^2 
\wedge dx^3
\eeqa

We are interested in the mirror of a IIB D3-brane, namely a D6 wrapping 
$0123x^1x^2x^3$. This shares many of the features of the D4-brane studied 
in previous sections. For instance, such D6-branes correspond to torsion 
classes in the 3-homology of the half-flat manifold, and hence carry no 
$\IZ$-valued charges. They are nevertheless stable and supersymmetric, 
since they are generalized calibrated, with respect to the form
\beqa
\label{caliD6(1)}
\Theta_{(6)}^{RR} & = & C_7\,-\,  e^{-\phi}\, \sqrt{-\det P[G]}\, dvol_{123}\wedge 
dx^1\wedge  dx^2\wedge  dx^3
\eeqa
which more explicitly reads
\beqa
\Theta_{(6)}^{RR} &=& \left ( -H^{3/4}\sqrt{(H^{-\oh})^7}+
(H^{-1}-1)\right) dvol_{123}\wedge dx^1\wedge  dx^2\wedge  dx^3=\nonumber \\
&=& -dvol_{123} \wedge dx^1 dx^2 dx^3
\eeqa

Given the nice geometric structure of half-flat manifolds as 
$SU(3)$-structure manifolds, it is natural to propose a characterization 
of generalized calibrated submanifolds in terms of the natural forms in 
the geometry, $J$ and $\Omega$. We indeed make such proposal, and provide 
evidence for it, in next section.

\section{Generalized calibrations in compactifications with fluxes and on 
$SU(3)$-structure manifolds}
\label{general}

\subsection{The proposal}
\label{sectpropo}

In this section we propose a characterization of generalized calibrated
submanifolds in flux and non-CY compactifications, in analogy with
standard calibrated submanifolds is standard Calabi-Yau compactifications\footnote{For extensive work on the relation between generalized calibrations and G-structures, mainly in the context of M-theory and fivebranes see \cite{martelli,dallagata} }.

Supersymmetric cycles in Calabi-Yau compactifications are nicely characterized 
in terms of the Kahler 2-form $J$ and the holomorphic 3-form $\Omega$ of 
the geometry. They fall in two classes, holomorphic cycles and special 
lagrangian 3-cycles, defined by familiar conditions, see below.

Supersymmetric compactifications with 3-form fluxes, and on non-CY geometries,
are generalizations of CY compactifications (see 
\cite{glmw,GranaMinasian} for a more extensive discussion). These 
backgrounds have a Killing spinor $\epsilon$, namely a globally defined 
spinor which is not covariantly constant with respect to the spin  
connection related to the metric, but is covariantly constant with respect 
to a connection with torsion. This implies that the structure group of 
the tangent bundle is $SU(3)$, hence the name $SU(3)$-structure 
compactifications. The invariant spinor may be used to construct a 
globally defined 2-form $J$ and 3-form $\Omega$, which in contrast with 
the CY case, are not closed. Their exterior derivatives are classified by 
the torsion classes of the connection, and are related to the non-metric 
part of the background (see e.g. \cite{glmw}).

The last statements are very suggestive of a relation with generalized 
calibrations. Indeed we suggest that generalized calibrated submanifolds 
in $SU(3)$-structure manifolds are characterized, in analogy with the 
Calabi-Yau case, using the 3-form $\Omega$ and a complexified version of 
$J$, namely $e^{J+iB}$. The geometrical structure suggests the 
definitions of\\
\textbf{i)} (generalized) holomorphic cycles, characterized by the conditions 
\beqa
\label{holomorphic}
\Omega\vert_\xi=0 \quad ; \quad e^{J+iB}\vert_\xi=vol_\xi
\eeqa\\
\textbf{ii)} (generalized) special lagrangian 3-cycles, characterized by
\beqa
\label{slag}
(J+iB)\vert_\xi=0 \quad ; \quad \pim\Omega\vert_\xi=0 \quad ;\quad 
\preal\Omega\vert_\xi = vol_\xi
\eeqa

The generalized structure of these calibrated cycles lies not in the 
conditions, but in the property that the forms $J+iB$, $\Omega$ are not 
closed. In order to write the corresponding generalized calibrations, it 
is convenient to factorize the volume term into a piece from the 
non-compact spacetime and a piece from the internal submanifold volume, 
which can be recast using (\ref{holomorphic}), (\ref{slag}). Denoting 
$\omega\vert_\xi=e^{J+iB}\vert_\xi$ for holomorphic and 
$\omega\vert_\xi=\preal\Omega\vert_\xi$ for special lagrangian 
submanifolds, the generalized calibration, for D-branes, may be written
\beqa
\Theta_{(Dp)}=\Theta_{CS} 
-e^{-\phi}\sqrt{-\det{P[G]}} \, \tilde{\Phi}_{0}\wedge \omega|_\xi
\eeqa
where $\Theta_{CS}$ contains the terms associated to CS couplings, and 
$\tilde{\Phi}_{0}$ refers to the non-compact piece of the world-volume.

This proposal could presumably be verified by using the expression of the 
supersymmetry variations of general supergravity backgrounds in terms of 
$\Omega$ and $e^{J+iB}$ \cite{GranaMinasian}. Notice also that the 
exchange of both forms under mirror symmetry \cite{GranaMinasian} suggest 
a mirror exchange of both kinds of branes, generalizing the familiar map 
for D-branes on Calabi-Yau spaces. We leave these very interesting 
exploration for future research. Here we limit ourselves to verifying the 
above proposal in a set of examples, obtained from (generalized 
holomorphic) D-branes in 3-form flux backgrounds, and (generalized 
special lagrangian) D6-branes in half-flat geometries (which for
suitable NSNS fluxes are related by mirror symmetry \cite{glmw}). Still 
this leads to interesting results, like the stabilization of D7-brane 
moduli by 3-form fluxes. It would be nice to demonstrate the above 
relation in more generality, and for other kinds of branes.

\subsection{The D3-brane and its mirror, revisited}

One first check of the proposal is to recover the results of section 
\ref{sectD6(1)} from the above prescription. The case of D3-branes in a 
general supersymmetric 3-form flux background is straightforward. They 
correspond to generalized holomorphic calibrated submanifolds, because 
they satisfy $\Omega|_\xi=0$ since D3-branes span just a point in the 
internal space. They are supersymmetric, and the generalized calibration 
contains a volume term related to the restriction of $e^{J+i\Bb}$, in a 
trivial way.

A more interesting case is provided by the generalized calibration 
structure of its mirror D6-brane, in the half-flat background 
(\ref{halfflatmetric}). 
It wraps a 3-submanifold of the internal space, which we claim is a 
generalized special lagrangian calibrated submanifold, according to our 
description above.
Indeed, using the complex orthogonal frames (\ref{frames}), the 2- and 
3-forms $J$ and $\Omega$ (\ref{jomega}) read
\beqa
\label{JOmeD6(1)}
J&=&-\frac{i}{2} \left ( e_1\wedge \ov{e_1}+e_2\wedge \ov{e_2}+e_3\wedge 
\ov{e_3}  \right )=\sum_{i=1}^{3}dy^i \wedge \left ( dx^i+g_{(x^i)} 
\right )\\
\Omega&=&e_1 \wedge e_2 \wedge e_3= \nonumber\\
&=&\left\{ H^{-3/4}(dx^1+g_{(x^1)})\wedge 
(dx^2+g_{(x^2)})\wedge(dx^3+g_{(x^3)}) 
- H^{1/4}dy^1\wedge dy^2 \wedge(dx^3+g_{(x^3)})\right.  \nonumber\\  && 
\left . -H^{1/4}dy^1 \wedge (dx^2+g_{(x^2)})\wedge dy^3- 
H^{1/4}(dx^1+g_{(x^1)}) \wedge dy^2 \wedge dy^3\right \} \nonumber \\
&+i&\left \{H^{-1/4}(dx^1+g_{(x^1)})\wedge(dx^2+g_{(x^2)})\wedge 
dy^3 +H^{-1/4} (dx^1+g_{(x^1)})\wedge dy^2 \wedge(dx^3+g_{(x^3)}) \right 
. \nonumber \\ 
&&+ \left. H^{-1/4}dy^1 \wedge 
(dx^2+g_{(x^2)})\wedge(dx^3+g_{(x^3)})-H^{3/4}dy^1\wedge dy^2 \wedge dy^3 
\right \}
\eeqa
We can readily check that the conditions (\ref{slag}) are automatically 
satisfied for the D6-brane along $x^1$, $x^2$, $x^3$, since all 
components of $J$ and $\pim \Omega$ have at least one leg along $dy^i$.

Moreover, according to our proposal, the D6-brane should be calibrated 
with respect to a generalized calibration, with the volume piece related 
to $\preal \Omega$. Indeed it is straightforward to check that the 
generalized calibration (\ref{caliD6(1)}) can be written
\beqa
\Theta_{(6)}^{RR}=C_{7}-e^{-\phi}\sqrt{\det 
P[G]_{0123}}dvol_{123}\wedge\preal \Omega
\eeqa

\subsection{D7 branes in the presence of $\NN=2$ $G_3$ fluxes.}
\label{sectD7-1}
Let us now turn to some new examples. Consider IIB theory on $T^6$ or a 
partially decompactified version thereof, in the background generated by 
the 3-form fluxes 
\beqa
\Hh_3&=&-dx^1dy^1dy^3+dx^2dy^2dy^3\nonumber\\
F_3&=&dx^1dy^1dx^3-dx^2dy^2dx^3
\eeqa
which are ISD and in fact $N=2$ supersymmetric for $\tau=i$. Using the 
complex coordinates $z^j=x^j+iy^j$, we have
\beqa
G_3\, =\frac{i}{2}\, \left(\, dz_1\, d\ov{z}_1\, dz_3 \, -\, 
dz_2\, d{\ov z}_2\, dz_3 \, \right )
\eeqa
The associated potentials, in a suitable gauge, read  
\beqa
\label{D7(1)2forms}
\Bb_2&=&-y^3 \ dx^1dy^1 + y^3 \ dx^2dy^2\nonumber\\
C_2&=& x^3 \ dx^1dy^1-x^3 \ dx^2dy^2
\eeqa
They generate a black 3-brane background
\beqa
ds^2&=&H^{-\oh}(ds^2_{4d})+H^{\oh}\left ( (d\vec{x})^2+(d\vec{y})^2 
\right )\nonumber \\
C_4&=&(H^{-1}-1)dvol_{0123}
\eeqa
Concerning the dual RR potentials $C_8, C_6$, and straightforward 
computation shows that they vanish in our particular background (see  
appendix for conventions on generalized field strengths, etc).

Let us consider a D7-brane spanning 0123 and the internal directions
$z_1$, $\ov{z}_1$, $z_2$, $\ov{z}_2$, and located at a fixed coordinate 
$z_3$. Since this brane is wrapped on a holomorphic 4-cycle of the 
underlying Calabi-Yau, it is expected to be supersymmetric. Indeed, in 
the remainder of this section, we show that it is generalized calibrated, 
and that it can be regarded as a generalized holomorphic brane, in the 
description of section \ref{sectpropo}.

Since the background contains non-zero RR fluxes and NSNS $\Bb$-field, 
the appropriate generalized calibration for this case is the one in 
section \ref{gensitu}. For our D7-brane the expression (\ref{Bgencali}) 
reads
\beqa
\label{D7(1)bruto}
\left . \Theta^{RR}_{(7)}\right |_{\xi}\, =\, i_{K}\left(\, C_{8}\, +\, 
C_6 \wedge \Bb_2\, +\, \oh C_4 \wedge \Bb_2 \wedge \Bb_2 \,+\cdots 
\,\right)\, +\, \left . e^{-\phi}\,
\sqrt{\det \left (P[g+\Bb] \right )}\, \tilde{\Phi}^{(7)}_{(0)}\right |_{\xi}
\eeqa
Note that the relative sign of the second term has changed with respect 
to that in former sections. The reason is that we consider a D7 wrapping 
the holomorphic cycle with volume form $dz^1 \wedge d\ov{z}^1\wedge dz^2 
\wedge d\ov{z}^2$. The volume form we are using in real coordinates $dx^1 
\wedge dx^2 \wedge dy^1 \wedge dy^2$, which has opposite orientation, 
hence yielding an extra sign.

The determinant of the pullback of $g+\Bb$, and the contribution from the 
Chern-Simons coupling, are easily computed to be
\beqa
\label{D7(1)DBI}
\sqrt{\det P[g+\Bb]} & = &H^{-1}\,(\, H\,+\,(y^3)^2\,) \nonumber \\
C_4 \wedge \Bb_2 \wedge \Bb_2 & = &- 2 H^{-1}\ (y^3)^2 dvol_{0123}\wedge 
dx^1 dy^1 dx^2 dy^2
\eeqa
We then have
\beqa
\label{D7(1)complete}
\left .\Theta_{(7)}^{RR}\right |_{D7}&=&\left ( 
-H^{-1}(y^3)^2+H^{-1}\left ( H+(y^3)^2 \right ) 
\right )dvol_{0123}\wedge dx^1 dy^1 dx^2 dy^2=\nonumber \\ &=& 
dvol_{0123}\wedge dx^1 dy^1 dx^2 dy^2
\eeqa
Hence the expression is closed and corresponds to a generalized 
calibration, and the D7-brane is generalized calibrated. Notice that this is 
so for any value of the D7-brane transverse coordinate $z_3$. In section 
\ref{fixing} we will describe an example where the D7-brane is 
generalized calibrated (i.e. supersymmetric) only at a fixed transverse 
position.

\medskip

Let us now describe the above generalized calibrated submanifold in terms 
of the conditions in section \ref{sectpropo}. We introduce a set of 
complex vielbeins 
\beqa
\label{D7(1)viel}
e_i=H^{1/4}(dx^i+i dy^i)
\eeqa
The pullback of the 3-form $\Omega=e_1\wedge e_2 \wedge e_3$ on the 
D7-brane is automatically zero. Hence, the D7-brane spans a generalized 
holomorphic calibrated 
4-submanifold.  Moreover, the generalized calibration should contain a 
volume piece associated to $e^{J+i\Bb}$, or rather its relevant piece for 
a 4-submanifold, $\frac 12(J+i\Bb)^2$. Indeed, using (\ref{D7(1)2forms}) 
and (\ref{D7(1)viel}), we have
\beqa
\label{D7(1)J}
J+i\Bb&=&-\frac{i}{2} \sum_i e_i\wedge \ov{e}_i+i\Bb = H^{\oh} 
\left ( dx^1 \wedge dy^1 + dx^2 \wedge dy^2+  dx^3 \wedge dy^3 \right 
)+\nonumber \\&& +i\left (-y^3dx^1\wedge dy^1 + y^3 dx^2 
\wedge dy^2 \right )\nonumber \\
(J+i\Bb)\wedge(J+i\Bb)|_{D7}&=&2 \left ( H - \Bb_{x_1 y_1 }\Bb_{x_2 
y_2}\right )dx^1\wedge  
dy^1\wedge  dx^2 \wedge dy^2=\nonumber \\ && 2\left ( H + (y^3)^2\right 
)dx^1\wedge  dy^1\wedge  dx^2\wedge  dy^2
\eeqa
This indeed reproduces (once the factor from the non-compact directions 
is considered) the volume piece of the generalized calibration 
(\ref{D7(1)DBI}), which can thus be recast as
\beqa
\label{D7(1)sofis}
\left. \Theta_{(7)}^{RR}\right |_{\xi}&=&\left. i_K\left ( \oh C_4 \wedge 
\Bb_2 
\wedge \Bb_2\right ) + e^{-\phi}\sqrt{\det P[G]_{0123} } \, \oh 
(J+i\Bb)\wedge (J+i\Bb)\tilde{\Phi}_{0} \right |_{\xi}
\quad \quad \nonumber 
\eeqa 
Let us consider the mirror version of this system, by carrying out three 
T-dualities along the $x^i$ coordinates. Using the results in section 
\ref{sectD6(1)}, the NSNS background in the IIA side is 
\beqa
ds^2 & = & H^{-\oh}\, \left(\, ds^2_{0123}\, +\, (\, dx^1+g_{(x^1)}\,)^2\, +\, 
(\,dx^2+g_{(x^2)}\,)^2\, +\, (dx^3)^2\, \right)\, +\nonumber\\
&& +\,H^{\oh}\, 
\left (\, (dy_1)^2+(dy_2)^2+(dy^3)^2 \, \right) \nonumber \\
e^{-\phi} & = & H^{3/4} 
\eeqa
with $B_2=0$ and $g_{(x^i)}\equiv -\Bb_{x^i y^{\a}}dy^\a$. In our case
\beqa
g_{(x^1)}&=&y^3 \ dy^1 \nonumber \\
g_{(x^2)}&=&-y^3 \ dy^2 \nonumber \\
g_{(x^3)}&=&0
\eeqa
Concerning the RR potentials, the only component relevant for our 
purposes below is $(C_7)_{0123 y^1 y^2 x^3}$ which, from our original 
configuration and using the T-duality formulas in the appendix, turns out 
to be:
\beqa
(C_7)_{0123 y^1 y^2 x^3}=-H^{-1}(y^3)^2 dvol_{0123} dy^1 dy^2 dx^3 
\eeqa
We are interested in describing the mirror of the original D7-brane, 
which is given by a D6-brane along the coordinates $0123$ and $y^1$, 
$y^2$, $x^3$.
Since there is no $B$-field present, the generalized calibration is provided by
(\ref{gencalibr}), namely
\beqa
\label{D6(2)cali}
\left . \Theta_{(6)}^{RR}\right |_{D6}=i_K (C_{7})+e^{-\phi}\sqrt{\det 
P[G]}dvol_{123}\wedge dy^1\wedge  dy^2 \wedge dx^3
\eeqa
where $P[G]$ is the pullback of the metric onto the full (spacetime) 
world-volume directions. Notice that this dual realization shows that 
expression (\ref{Bgencali}) is the generalized calibration to be used in 
the original IIB case.

By considering the pullbacked metric
\beqa
\left. ds^2\right |_{D6}=H^{-\oh} \left ( ds^2_{0123} +(y^3)^{\,2} \ 
(dy^1)^2+(y^3)^{\,2} \ 
(dy^2)^2+(dx^3)^2 \right )+ H^{\oh}\left ((dy^1)^2 +(dy^2)^2\right )
\quad
\eeqa
whose determinant square root reads
\beqa
\sqrt{-\det P[G]}=H^{-7/4}(H+(y^3)^2)
\eeqa 
we get the result
\beqa
\left . \Theta^{RR}_6\right |_{D6}&=&\left ( -H^{-1}\, (y^3)^{\,2} 
+H^{-1}\left (H+(y^3)^{\,2} \right 
)\right )dvol_{0123}\wedge dy^1 dy^2 dx^3 \nonumber \\ 
&=&dvol_{0123}\wedge dy^1 dy^2 dx^3
\eeqa
Hence the D6-brane is calibrated in this half-flat manifold, as expected 
from the mirror image. Notice that the result obtained in IIA with the 
generalized calibration (\ref{gencalibr}) reproduces the mirror result in 
IIB with the generalized calibration (\ref{Bgencali}). This lends 
additional support for the proposal to include NSNS fields (and 
world-volume gauge couplings) in section \ref{gensitu}.

Finally, we would like to describe this D6-brane as a generalized special 
lagrangian calibration. Introducing a complex orthonormal frame, and 
defining $J$ and $\Omega$ as usual, we again find that the two conditions 
in (\ref{slag}) are automatically satisfied.

Hence the 3-cycle satisfies the generalized special lagrangian 
calibration conditions. Moreover, the pullback of $\preal\Omega$ should 
provide the volume piece in the generalized calibration. Indeed, we have
\beqa
\left. \preal\Omega\right|_{\xi}=-H^{-3/4}(H+(y^3)^2)\,
dy^1\wedge dy^2\wedge dx^3
\eeqa 
so that (\ref{D6(2)cali}) may be rewritten as
\beqa
\Theta_{(6)}^{RR}=i_K(C_{7})-e^{-\phi}\sqrt{\det 
P[G]_{0123}}\,dvol_{123}\wedge\preal(\Omega)
\eeqa

We hope these examples suffice to illustrate how the proposal in section 
\ref{sectpropo} can be used to characterize supersymmetric wrapped branes in
flux backgrounds and half-flat manifold. We also expect it to have some 
more general validity in the complete class of $SU(3)$-structure 
manifolds. Unfortunately, lack of explicit examples makes it difficult to 
carry further tests of this stronger form of the proposal.

\subsection{Generalized calibrations and fixing D-brane moduli}
\label{fixing}

In this section we describe an example of a brane which is generalized 
calibrated only at a particular position of its transverse space. In the 
type IIB picture of branes in flux backgrounds, this reproduces how 
fluxes fix moduli of D-branes wrapped on non-rigid cycles of the 
underlying Calabi-Yau. 

Consider a type IIB $(2,1)$ and primitive $G_3$ flux background
\beqa
G_3=dz_1dz_2d\ov{z}_3
\eeqa
In real coordinates, for $\tau=i$, the corresponding RR and NSNS 3-form 
fluxes, and 2-form potentials, read
\beqa
F_3 & = & dx^1dx^2dx^3\, +\, dx^1dy^2dy^3\, -\, dy^1dy^2dx^3\, +\, 
dy^1dx^2dy^3\nonumber \\
\Hh_3 & = & dx^1dx^2dy^3\, -\, dx^1dy^2dx^3\, -\, dy^1dx^2dx^3\, -\, 
dy^1dy^2dy^3\nonumber \\
C_2 & = & x^3dx^1dx^2\,  +\, y^3dx^1dy^2\, -\, x^3dy^1dy^2\, +\, 
y^3dy^1dx^2\nonumber \\
\Bb_2 & = & y^3dx^1dx^2\, -\, x^3dx^1dy^2\, -\, x^3dy^1dx^2\, -\, 
y^3dy^1dy^2
\eeqa
As usual, these generate a black 3-brane background of the form:
\beqa
ds^2&=&H^{-\oh}(ds^2_{0123})+H^{\oh}\left ( 
(d\vec{x})^2+(d\vec{y})^2\right )\nonumber \\
C_4&=&H^{-1}dvol_{0123}
\eeqa
(where we dropped the constant term in $C_4$ for simplicity).

Consider a D7-brane spanning 0123 and $z_1$, ${\ov z}_1$, $z_2$, ${\ov 
z}_2$, and located at a point in the $z_3$ coordinate. Our purpose is to 
understand the generalized calibrated structure of such D7-branes.

In principle, one can work as in previous sections, with the expression 
(\ref{D7(1)bruto}) for the calibrating form for a D7-brane in a 3-form flux 
background. Instead, we follow the simpler approach introduced in 
previous section, of describing the generalized calibration conditions 
directly from $J$ and $\Omega\,$ \footnote{We have indeed verified that 
the explicit construction of the calibrating form leads to equivalent 
results.}. The D7-brane probe wraps a generalized holomorphic 4-cycle of 
the internal space, corresponding to the fact that it satisfies 
$\Omega|_\xi=0$. Moreover, it must be generalized calibrated when it 
minimizes its action
\beqa
\label{D7(3)cali}
\Theta_{(7)}^{RR}&=&
i_K\left ( C_8 + C_6\wedge \Bb_2 +\oh C_4 \wedge 
\Bb_2 \wedge \Bb_2+\cdots \right )+\nonumber \\&+& e^{-\phi}\sqrt{\det 
P[G]_{123}}H^{-1/4}dvol_{123}\wedge \oh\left .  (J+i\Bb)\wedge 
(J+i\Bb)\right |_{\xi}
\eeqa
where again the sign of the second term has changed with respect to the 
original formulation due to orientation conventions.

We have
\beqa
J+i\Bb=&-&\frac{i}{2} \sum_i e_i\wedge \ov{e}_i+i\Bb=H^{\oh} \left ( dx^1 
\wedge 
dy^1 + dx^2 \wedge dy^2+  dx^3 \wedge dy^3 \right )+\nonumber 
\\&+&i(y^3dx^1\wedge dx^2\, -\, 
x^3dx^1\wedge dy^2\, -\, x^3dy^1\wedge dx^2\, -\, 
y^3dy^1\wedge dy^2)
\eeqa
which leads to
\beqa
\left. (J+i\Bb)\wedge (J+i\Bb)\right|_{\xi}=2\left ( H+|z_3|^2\right )
\eeqa
Concerning the Chern-Simons part of the calibration, it involves the 
$C_6$ and $C_8$ RR dual gauge potentials. However, one can easily compute that they again vanish in our case, so that the only relevant Chern-Simons coupling is:
\beqa
C_4\wedge \Bb_2 \wedge \Bb_2&=&2H^{-1}|z_3|^2dvol_{0123}\wedge dx^1\wedge 
dy^1\wedge dx^2\wedge dy^2
\eeqa
In total, expression (\ref{D7(3)cali}) reads
\beqa
\left. \Theta_{(7)}^{RR}\right |_{\xi} &=&
\left ( H^{-1}|z_3|^2+H^{-1}(H+|z_3|^2)\right )dvol_{0123}\wedge dx^1\wedge 
dy^1\wedge dx^2\wedge dy^2\, = \nonumber \\
&=&\left( \, 1+2H^{-1}|z_3|^2\,\right )\, dvol_{0123}\wedge dx^1\wedge 
dy^1\wedge dx^2\wedge dy^2
\eeqa
Notice that this expression is not closed, in agreement with our comments 
that expression (\ref{Bgencali}) in general provides the action for a 
D-brane configuration rather than the generalized calibration. The action 
is minimized for $z_3=0$, where it takes the value
\beqa
\left. 
\Theta_{(7)}^{RR}\right |_{\xi} &=& dvol_{0123}\wedge dx^1\wedge 
dy^1\wedge dx^2\wedge dy^2
\eeqa
This agrees with the restriction to the world-volume of a closed 
generalized calibrated form. 
Hence, interestingly the D7 is generalized 
calibrated only if it is 
located at $z_3=0$ in transverse space. This simply means that in the 
system at hand there  is a non-trivial superpotential for the D7-brane 
geometric moduli $\varphi_3$ parameterizing its position in $z_3$, such 
that the only supersymmetric minimum is at $\varphi_3=0$.

This fits nicely with a similar result obtained from alternative 
approaches, where the effect of fluxes on D7-branes is computed directly. 
Indeed, a quadratic superpotential arises for D7-brane moduli in certain 
supersymmetric fluxes, as shown in \cite{ciu2} from the D7-brane 
world-volume perspective, and in \cite{gktt} from the F-theory 
perspective. This result is important in that it modifies the infrared 
dynamics of D7-brane gauge theories in flux backgrounds. In particular, 
as already mentioned in \cite{kklt}, flux effects generate masses for 
D7-brane matter, even in supersymmetric situations, hence leading to 
non-perturbative superpotentials from gaugino condensation, which provide 
a source of stabilization for Kahler moduli. We hope that the techniques 
developed here (as well as in complementary approaches) help in rendering 
this effect tractable, so that it can be quantitatively included in 
further discussions of the construction of models with full moduli  
stabilization.

\section{Model building with NS-branes}
\label{building}

We would like to mention another possible application of branes wrapped on 
generalized calibrated submanifolds. In particular, one can take advantage 
of the fact that such configurations can be supersymmetric even if the 
branes carry no charges. This can be employed in model building since it 
allows to avoid the (sometimes very constraining) tadpole cancellation 
conditions present in the familiar situations with charged branes. 
It may be argued that branes carrying no charge in homology are 
unable to lead to chiral fermions; we will see below how net chirality may 
be achieved even for branes in homologically trivial cycles.

In addition, it allows model building with somewhat unfamiliar branes, for 
which no orientifold planes exist, like NS5-branes \footnote{The analogs 
of orientifold planes for NS5-branes can be constructed via duality (e.g. as 
S-dual of O5-planes in type IIB theory). However, this description is not 
explicit enough to be practical in model building.}. 
This is particularly interesting, since NS5-branes are a key ingredient in 
the certain configurations of branes leading to chiral gauge theories, 
namely brane boxes \cite{bboxes} and brane diamonds \cite{bdiamonds}. 
Moreover, this source of chirality in string theory has not been exploited 
in model building, due to the difficulties in discussing NS-branes in 
compact examples. In the present section, our purpose is to illustrate 
the construction of chiral 4d gauge sectors of NS- and D-branes wrapped 
on generalized calibrated submanifolds. 

\medskip

Consider the background AdS$_5\times$ S$^5$, with $N$ units of 
5-form field strength flux through the S$^5$. Let us parametrize AdS$_5$ 
by the coordinates $x^\mu$, $\mu=0,\ldots,3$ and $r$, and parametrize 
S$^5$ as the unit sphere in $\IR^6$ parametrized by $x^m$, $m=4,\ldots, 9$. 
The complete metric of the background is
\beqa
ds^2 & = & H(r)^{-1/2} \, \eta_{\mu\nu} \, dx^\mu \, dx^\nu \, + \, 
H(r)^{1/2} \, [\, (dx^4)^2 + \ldots + (dx^9)^2 \,]
\eeqa
as discussed in section \ref{examples}.

We would like to consider a configuration of wrapped calibrated branes in this geometry. 
Consider a NS5-brane spanning 012, the 
radial direction $r$, and the maximal $S^2$ at $x^7=x^8=x^9=0$ (hence 
described as the unit $S^2$ in the $\IR^3$ parametrized by $x^4$, $x^5$, 
$x^6$). The world-volume is embedded in an AdS$_4\times S^2$ geometry inside 
AdS$_5\times S^5$. Consider a second NS5-brane (denoted NS'-brane) 
spanning 012, the radial direction $r$, and the maximal $S^2$ at 
$x^5=x^6=x^7=0$ (hence 
described as the unit $S^2$ in the $\IR^3$ parametrized by $x^4$, $x^8$, 
$x^9$). The world-volume is embedded in an AdS$_4\times S^2$ geometry inside 
AdS$_5\times S^5$. These two $S^2$'s intersect at two points, 
corresponding to $x^4=\pm 1$ in the $S^5$.

Finally, introduce a set of $n$ D5-branes spanning 012, the radial 
direction $r$, and the 
maximal $S^2$ at $x^6=x^7=x^9=0$ (hence described as the unit $S^2$ in 
the $\IR^3$ parametrized by $x^4$, $x^5$, $x^8$). The world-volume is 
embedded in a AdS$_4\times S^2$ geometry inside AdS$_5\times S^5$. The 
complete set of branes and their orientation is sketched in the following 
table

\begin{center}

\begin{tabular}{ccccccccccc}
& 0 & 1 & 2 & 3 & 4 & 5 & 6 & 7 & 8 & 9 \\
\hline
NS & $-$ & $-$ & $-$ & x & $-$ & $-$ & $-$ & x & x & x \\
NS' & $-$ & $-$ & $-$ & x & $-$ & x & x & x & $-$ & $-$ \\
D5 & $-$ & $-$ & $-$ & x & $-$ & $-$ & x & x & $-$ & x 
\end{tabular}
\end{center}
where $-$ denotes a direction spanned by the brane volume, and x denotes 
a direction transverse to it. Also $r$ should be regarded as the radial 
direction in the $\IR^6$ parametrized by $x^m$.

\begin{figure}
\begin{center}
\centering
\epsfysize=3.5cm
\leavevmode
\epsfbox{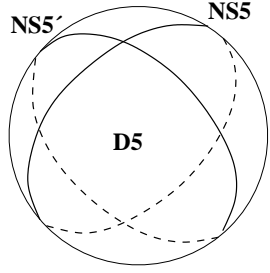}
\end{center}
\caption[]{\small The D5-branes wrap an $S^2$ in $S^5$, which is cut in four 
quadrants by its intersection with the NS- and NS'-brane $S^2$'s.}
\label{box}
\end{figure}

All branes are supersymmetric, since they correspond to calibrations of 
the kind considered in section \ref{exotic}.
Notice that all branes span a common AdS$_4$, and that in the $S^5$, the 
geometry is as follows. Each NS5-brane intersects with the D5-brane in a 
(different) circle $S^1$ in $S^5$. The D5-brane $S^2$ is therefore cut in 
four quadrants by these two $S^1$'s, like two orthogonal meridians, which 
moreover touch at the two poles $x^4=\pm 1$, as shown in figure 
\ref{box}.

Moreover, it is possible to verify that the configuration preserves 1/8 of the 
supersymmetries of the background, namely four supercharges, or 4d $\NN=1$ 
in the common AdS$_4$ dimensions. Moreover, notice that although we have a 
compactification on $S^5$ and a non-trivial set of branes which are stable 
and supersymmetric, there is no need to introduce orientifold planes, 
since all branes wrap homologically trivial cycles, and therefore carry no 
charge.

Interestingly, this set of branes leads to a non-trivial chiral 4d gauge 
theory in the common AdS$_4$. One simple way to notice it it to realize 
that the local geometry near the intersections of the three branes is 
exactly that arising in brane box models \cite{bboxes}, where D5-branes 
are suspended among intersecting NS-branes. Indeed, the complete 
configuration can be regarded as two sets of intersecting NS-branes, with 
D5-branes suspended between them.
\begin{figure}
\begin{center}
\centering
\epsfysize=3.5cm
\leavevmode
\epsfbox{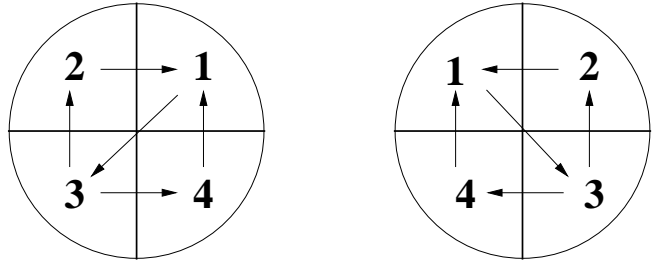}
\end{center}
\caption[]{\small North and South pole views of the $S^2$ spanned by the D5-branes. The numbers label the gauge groups arising from D5-branes in each quadrant, while the arrows denote chiral multiplets arising from the brane box intersections. Notice that the gluing of the two half-spheres makes some of the arrows in the picture redundant.}
\label{arrow1}
\end{figure}
Since chirality arises from the local intersections between branes, one may use the local features of brane box models to obtain the gauge 
group and matter content on the chiral gauge theory on the D5-branes.
Since the D5-brane $S^2$ is cut in four pieces by the NS5-branes, the 
gauge group is $U(n)^4$. The chiral multiplet content can be described in 
terms of arrows stretching between adjacent branes, in a particular way 
\cite{bboxes}. The set of arrows for our configuration is shown in 
figure \ref{arrow1}, and leads to the chiral multiplet content
\begin{center}
\begin{tabular}{cccc}
$U(n)_1$ & $U(n)_2$ & $U(n)_3$ & $U(n)_4$ \\
$\fund$ & & $\antifund$ & \\
$\fund$ & & $\antifund$ & \\
$\antifund$ & $\fund$ & & \\
$\antifund$ & & & $\fund$  \\
& $\antifund$ & $\fund$ &  \\
& & $\fund$ & $\antifund$ 
\end{tabular}
\end{center}
An important difference with respect to \cite{bboxes} is that in our 
model the volumes of the NS-branes are compact, hence they lead to 
dynamical degrees of freedom in AdS$_4$. It is difficult to determine them 
in detail, but they clearly lead to matter uncharged under the D5-brane 
gauge group, and will not be further discussed. We simply point out that, 
due to compactness of the NS-branes, the $U(1)$ gauge factors on the 
D5-branes presumably remain massless, in contrast with standard brane box 
models.

\medskip

\begin{figure}
\begin{center}
\centering
\epsfysize=3.5cm
\leavevmode
\epsfbox{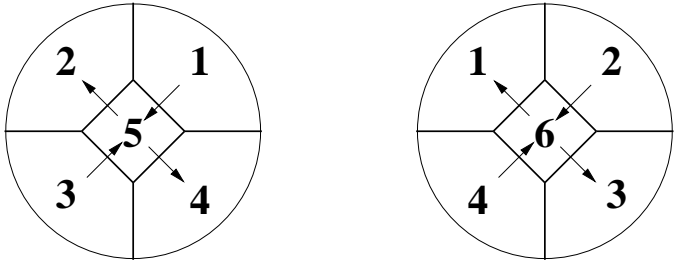}
\end{center}
\caption[]{\small North and South pole view of the calibrated branes in the brane diamond configuration.}
\label{diamond}
\end{figure}

A more precise version of the brane box models was introduced in 
\cite{bdiamonds}, where each intersection between NS-branes is resolved 
into a smooth recombination, leading to a diamond-shaped region, where 
additional D5-branes can be suspended. In our present setup, the brane 
diamond configuration is easily obtained by considering our NS-branes to 
span the holomorphic 2-cycle $\Sigma$ defined by $z_1z_2=\epsilon$, with 
$z_1$, $z_2$ complex coordinates in the 45 and 89 two-planes 
respectively. An NS5-brane spanning AdS$_4\times\Sigma$ is 
calibrated, in analogy with section \ref{hermitian}. The resulting brane 
diamond picture is shown in figure \ref{diamond}, along with the 
corresponding 
arrows, from \cite{bdiamonds}. The final 4d chiral gauge theory in AdS$_4$ 
is
\beqa
& U(n)_1\times U(n)_2\times U(n)_3\times U(n)_4\times U(n)_5\times U(n)_6 
& \nonumber \\
& (\fund_1,\antifund_5) + (\antifund_1,\fund_6) + 
(\fund_3,\antifund_5) + (\antifund_3,\fund_6) + & \nonumber \\
& + (\antifund_2,\fund_5)+(\fund_2,\antifund_6)
+ (\antifund_4,\fund_5)+(\fund_4,\antifund_6) &
\eeqa

Clearly, generalizations of the above setup are possible. Introducing a 
more general network of intersecting NS-branes (or alternatively 
NS5-branes in more general holomorphic 2-cycles, which are 
similarly calibrated and supersymmetric) leads to a larger and richer class of 4d chiral gauge theories which can be engineered in this way. It 
would be interesting to explore the possible phenomenological applications 
of this kind of construction.

In this respect it is interesting to point out the possibility of somewhat 
relaxing the condition to preserve a common supersymmetry, and allow 
cycles which are calibrated with different phases. Intuitively, this implies that the angles between NS-branes at their 
intersections do not satisfy the $SU(2)$ condition $\theta_1\pm 
\theta_2=0$. In compactifications to flat space, such configurations would 
lead to instabilities against decay to a recombined branes, whose volume 
is smaller than the intersecting set. In compactifications to AdS space, 
the potential unstable mode is still above the BF bound, and hence no 
instability is found, at least for certain range of non-supersymmetric angles \footnote{This follow from the analogous flavoured AdS/CFT analysis in \cite{kk}, where the inter-brane mode between $SU(2)$ rotated branes corresponded to a tachyon finitely above the Breitenlohner-Freedman bound. Hence, for small enough AdS radius, small deviations from the $SU(2)$ condition do not lead to instabilities.}.

\medskip

Another interesting further issue is the possibility of obtaining 
theories with 4d gravity in the present setup. A simple possibility would 
be to compactify the direction $x^3$ of AdS$_5$, by making it periodic. 
Unfortunately, its prefactor in the metric makes the circle non-compact 
at infinity, rendering the gravitational interaction five-dimensional at 
long distances. Another possibility would be to recover 4d gravity from 
gravity localization, as in \cite{kr}, although this does not occur for 
the Poincare slicing we are exploiting. A less exotic possibility would 
be embedding the above setups in a global compactification, 
with a local region of the form AdS$_5\times S^5$, for instance a la 
Verlinde \cite{verlinde}, or in the way the local Klebanov-Strassler 
throat \cite{ks} is embedded in global compactifications \cite{gkp}. 
Since our gauge theory engineering is local, it  is valid in any such 
setup. We leave their detailed discussion for future research.

\section{Final comments}
\label{fcomments}

In this paper we have described generalized calibrations, and provided examples 
extending previous generalized calibrated submanifolds, in diverse 
backgrounds. We have applied this tool to the understanding of 
supersymmetric wrapped branes in flux compactifications and 
compactifications on $SU(3)$-structure manifolds, with several explicit 
examples, illustrating diverse phenomena.
For instance, the use of generalized calibrations allows a neat 
understanding of the supersymmetry of branes which carry no charges (or at 
least no $\IZ$-valued charges). Also, we have exploited generalized 
calibrations to show that certain supersymmetric fluxes stabilize D7-brane 
geometric moduli, via a flux-induced superpotential. This effect is 
important in the generation of non-perturbative superpotentials from 
strong dynamics of the D7-brane gauge theory, since the flux modifies the 
infrared matter content of the theory. Hence, it is crucial in the 
recently proposed mechanisms to stabilize Kahler moduli in flux 
compactifications \cite{kklt} (see \cite{ciu2,gktt} for related 
discussions). In addition, it is related to the interesting question of
moduli spaces of generalized calibrated D-branes.

We have also provided a simple characterization of generalized 
calibrated submanifolds, in terms of the 2- and 3-forms $J$ and $\Omega$ 
present in any $SU(3)$-structure compactification. We have verified the 
proposal in several examples of flux compactifications and 
compactifications on half-flat manifolds. It would be interesting to find 
supporting evidence beyond this class. Presumably, the more complete 
understanding of the whole class of $SU(3)$-structure manifolds in 
\cite{GranaMinasian} will lead to progress in this direction.

Finally, we have suggested that generalized calibrations can also be interesting beyond the above project. In particular, we have offered a new class of compactifications with chiral gauge sectors, based on generalized calibrated brane box configurations. This illustrates the new model building possibilities allowed by generalized calibrations.

We hope much progress in turning generalized calibrations into a central tool in
the understanding of string theory in general supersymmetric backgrounds.

\centerline{\bf Acknowledgements}

We thank K. Behrndt, M. Cveti\u{c}, T. Z. Husain, and T. Ort\'{\i}n for useful conversations. We specially thank P. G. C\'amara and L. E. Ib\'a\~nez for discussion on related topics, and S. Kachru and S. Trivedi for communications on \cite{gktt} prior to publication. A.M.U. thanks M.~Gonz\'alez for kind encouragement and  support. J.G.C. thanks M. P\'erez 
for her patience and affection. This work has been partially supported by 
CICYT (Spain). The research of J.G.C. is supported by the Ministerio de 
Educaci\'on, Cultura y Deporte through a FPU grant.

\newpage

\appendix

\section{Conventions and useful formulae}
\label{app}

For carrying out T-duality transformations in the presence of NSNS 3-form 
flux, it is useful to introduce the notation \cite{kstt}
\beqa
\label{defiKSTT}
[F_{n(x)}]&=&[F_n]_{x i_1...i_{n-1}} \, dx^{i_1}\ldots dx^{i_{n-1}} 
\nonumber\\
g_{(x)}&=&\frac{g_{x\a}}{g_{xx}}dx^\a \nonumber\\
\om_{(x)}&=&-d g_{(x)}
\eeqa

The rules for T-dualization of NSNS and RR fields are taken from 
\cite{Ortinbook}, see also \cite{myers}. They are:
\begin{itemize}
\item[$\bullet$] For NSNS fields:
\beqa
\tilde{G}_{xx} &=& \frac{1}{G_{xx}}\nonumber \\
\tilde{G}_{\a \b} &=& G_{\a \b}
- \frac{1}{G_{ xx}}\left ( G_{x \a} G_{x\b}
- B_{x\a} B_{x\b}\right)\nonumber\\
\tilde{G}_{x\a} &=&-\frac{B_{x\a}}{G_{xx}}\nonumber \\
\tilde{B}_{ \a \b}&=&B_{ \a \b}
-\frac{1}{G_{xx}}\left (G_{x\a} B_{x\b}-B_{x\a}
G_{x\b}\right )\nonumber\\
\tilde{B}_{x\a} &=&-\frac{G_{x \a}}{ G_{ xx}}\nonumber\\
e^{\tilde{\phi}} &=&\, \frac{e^{\phi} }{\sqrt{G_{xx}}}
\eeqa
where we call $x$ the T-duality coordinate. These expressions are valid 
for going from IIB to IIA or viceversa, and this is the reason why we  
have not kept the same notation as in the main text for distinguishing
IIB and IIA NSNS fields. We distinguish fields in one side and in the 
other by a tilde. The same applies for RR potentials.
\item[$\bullet$] For RR gauge potentials:
\beqa
\tilde{C}^{(n)}_{\a\cdots \b\gamma x}&=&C^{(n-1)}_{\a\cdots 
\b\gamma}-C^{(n-1)}_{[\a\cdots\b| x}\frac{G_{|\gamma]x}}{G_{xx}}\nonumber\\
\tilde{C}^{(n)}_{\a\cdots \b\gamma\delta}&=&C^{(n+1)}_{\a\cdots 
\b\gamma\delta x}+C^{(n-1)}_{[\a \cdots \b \gamma} 
B_{\delta]x}+C^{(n-1)}_{[\a\cdots 
\b|x}\frac{B_{|\gamma|x}G_{|\delta]x}}{G_{xx}}
\eeqa
\end{itemize}

We use a \textit{mostly plus} metric, and the ordinary expression for 
Hodge dualization in a $d$-dimensional space, namely:
\beqa
*(dx^{\a_1}\wedge \cdots\wedge dx^{\a_n} )=\frac{1}{p!}\sqrt{-\det G}\,
(g^{\a_1\b_1})\cdots (g^{\a_n\b_n})\e_{\b_1 \cdots \b_n \b_{n+1}\cdots 
\b_d}dx^{\b_{n+1}}\wedge \cdots\wedge dx^{\b_d}
\eeqa
The convention for the Levi-Civita tensor (with real indices) is 
$\e_{0123x^1y^1x^2y^2x^3y^3}=+1$.

We use the generalized field strength in the IIB supergravity 
theory. For those we follow the conventions in \cite{GHT} (with a shift $\Bb\longrightarrow -\Bb$), namely:
\beqa
\tilde{F}_{(2n+1)}=dC_{2n}+\Hh_{3}\wedge C_{(2n-2)}
\eeqa
which fulfill $\tilde{F}_5=*\tilde{F}_5$, $\tilde{F}_7=*\tilde{F}_3$ and 
$\tilde{F}_9=*\tilde{F}_1$. Taking into account that in our backgrounds 
$C_0=0$, we end up with the following equations for the dual fields $C_6$ 
and $C_8$:
\beqa
dC_6&=&-\Hh_3\wedge C_4 + *F_3 \nonumber \\
dC_8&=&-\Hh_3\wedge C_6
\eeqa
with $F_3=\tilde{F}_3=dC_2$ in the absence of axion. The expression for 
the dual gauge potentials is to be obtained from these two equations. 

Finally, as already pointed out in the main text, concerning the 
orientation for the internal holomorphic 4-cycles wrapped by the probe D7 
branes in sections \ref{sectD7-1} and \ref{fixing}, we consider the 
volume form $dz^1 \wedge d\ov{z}^1 \wedge dz^2 \wedge d\ov{z}^2$ with 
positive orientation. As a result, the volume form used in our 
calibrations $dx^1 \wedge dy^1 \wedge dx^2\wedge  dy^2$ has negative 
orientation.

%%%%%%%%%%%%%%%%%%%%%%%%%%%%%%%%%%%%%%%%%%%%%%%%%%%%%%%%%%%%%%%%%%%%%%%

\end{document}